\documentclass{article}

\usepackage{arxiv}

\usepackage[utf8]{inputenc} 
\usepackage[T1]{fontenc}    
\usepackage{hyperref}       
\usepackage{url}            
\usepackage{booktabs}       
\usepackage{amsfonts}       
\usepackage{nicefrac}       
\usepackage{microtype}      
\usepackage{lipsum}
\usepackage{graphicx}
\graphicspath{ {./images/} }
\usepackage{graphicx}%
\usepackage{multirow}%
\usepackage{amsmath,amssymb,amsfonts}%
\usepackage{amsthm}%
\usepackage{mathrsfs}%
\usepackage[title]{appendix}%
\usepackage{xcolor}%
\usepackage{textcomp}%
\usepackage{manyfoot}%
\usepackage{booktabs}%
\usepackage{algorithm}%
\usepackage{algorithmicx}%
\usepackage{algpseudocode}%
\usepackage{listings}%
\usepackage{url} 
\usepackage{hyperref}
\usepackage{orcidlink}
\usepackage{natbib}
\usepackage[singlelinecheck=false,labelfont=bf,font={small,stretch=1}]{caption}
\bibpunct{(}{)}{;}{a}{}{,}
\theoremstyle{thmstyleone}%
\theoremstyle{thmstyletwo}%
\theoremstyle{thmstylethree}%
\newtheorem{definition}{Definition}%
\title{A temporal graph model to study the dynamics of collective behavior and performance in team sports: an application to basketball}

\author{
 Quentin Bourgeais \\
  CETAPS UR 3832, Université Rouen Normandie\\
  F-76000 Rouen, France\\
 \And
 Eric Sanlaville \\
  LITIS UR 4108, Université Le Havre Normandie\\
  F-76600 Le Havre, France\\
  \And
 Rodolphe Charrier \\
  LITIS UR 4108, Université Le Havre Normandie\\
  F-76600 Le Havre, France\\
  \And
 Ludovic Seifert \\
  CETAPS UR 3832, Université Rouen Normandie\\
  F-76000 Rouen, France\\
}

\begin{document}

\maketitle

\begin{abstract}
In this study, a temporal graph model is designed to model the behavior of collective sports teams based on the networks of player interactions. The main motivation for the model is to integrate the temporal dimension into the analysis of players’ passing networks in order to gain deeper insights into the dynamics of system behavior, particularly how a system exploits the degeneracy property to self-regulate. First, the temporal graph model and the entropy measures used to assess the complexity of the dynamics of the network structure are introduced and illustrated. Second, an experiment using basketball data is conducted to investigate the relationship between the complexity level and team performance. This is accomplished by examining the correlations between the entropy measures in a team’s behavior and the team’s final performance, as well as the link between the relative score compared to that of the opponent and the entropy in the team’s behavior. Results indicate positive correlations between entropy measures and final team performance, and threshold values of relative score associated with changes in team behavior -- thereby revealing common and unique team signatures. From a complexity science perspective, the model proves useful for identifying key performance factors in team sports and for studying the effects of given constraints on the exploitation of degeneracy to organize team behavior through various network structures. Future research can easily extend the model and apply it to other types of social networks.
\end{abstract}


\newpage
\section{Introduction}\label{sec1}

In collective sports, a team can be defined as a group of players coordinating themselves to achieve a common goal. The complex systems approach has been proposed as a relevant paradigm for understanding team performance: the system (i.e., a team) consists of structurally and functionally heterogeneous components (i.e., players) that interact with varying intensities and across different spatiotemporal scales \citep{Balague2013}. Any in-depth understanding of a complex system must rely on system-level descriptions because any approach focusing only on the system entities would miss a fundamental ingredient of any complex system -- that is, the rich pattern of nonlinear interactions between the system components. Within this paradigm, networks have emerged as a reference modeling tool for complex systems \citep{Battiston2020}. Indeed, the assumption that studying only system components will not lead to a complete understanding of the system implies that the analysis of collective sports should emphasize interactions between players rather than the individual player actions. Collective sports teams can thus be seen as social networks of individuals in interaction \citep{newman2003structure}, and network analysis, through the Social Network Analysis (SNA) framework, has been introduced to collective sports analyses \citep{korte2019network}. Although representing a system as a graph is an easy way to get an overview of a system, this is usually done without accounting for the temporal dimension of the interactions \citep{holme2012temporal}. Yet, according to \citet{holme2012temporal}, the timing of interactions can have a major impact, and it might therefore be useful to consider the temporal dimension by modeling the system as a temporal graph. This holds true for collective sports analyses, as identified by \citet{Ramos2018}: a static network obtained by aggregating all interactions that occurred during an entire game may conceal important information and induce errors in certain metrics. Yet, thus far, the use of temporal graphs in collective sports has yet not been explored to any significant extent. In particular, the concept of \emph{attacking play} (or \emph{possession}, which can be defined as the tactical situation when one team is in possession of the ball moving toward the opponent’s target/goal in order to score) needs to be taken into account, which is possible through the use of temporal graphs \citep{Ramos2018}. This concept should be all the more relevant in sports in which numerous possessions follow one another in quick succession, such as in basketball or handball. Therefore, the analytical framework of the present study is defined as follows: the system is the offensive team in possession of the ball, its function is to score points, and its structure is the network of interactions between the players during the possession.

The goal of the present work is to propose a novel approach to modeling player interactions by integrating the temporal dimension and then to use the model to analyze the link between the interaction network and team performance in basketball. More precisely, it is an attempt to answer 2 questions: Is the level of complexity in the structure of a basketball team’s interaction network linked to the performance of this team (i.e., the final score of the game)? and Does the current performance of the team (i.e., the current difference in points with the opponent) affect the complexity level in this team’s interaction network structure? In this work, a collective sports team is considered as a complex system and it is hypothesized that the behavior of the system (i.e., the team’s behavior) emerges from the interactions between its components (i.e., the players) as shaped by the surrounding constraints (e.g., defense of the opposite team, game plan, score). As these constraints are constantly evolving, the system needs to produce adaptive reorganizations of its behavior and can be considered as a \emph{complex adaptive system}. Therefore, using concepts of coordination dynamics and ecological dynamics in collective sports sciences can help to understand which constraints lead to which reorganizations \citep{Balague2013}. The present work is precisely set within the ecological dynamics framework, which is based on the complexity science paradigm and predicated on: (1) the theory of constraints on dynamical systems, (2) ecological psychology, and (3) a complex systems approach in neurobiology \citep{Seifert2017}. 

Here, a fundamental concept of the ecological dynamics framework is mobilized: \emph{degeneracy}. In (neuro-)biology, degeneracy (or the principle of functional equivalence) is “the ability of elements that are structurally different to perform the same function or yield the same output” \citep{edelman2001degeneracy}. In other words, a system’s degeneracy property allows the same function to be performed by several different structures. According to \citet{edelman2001degeneracy}, this property appears at all levels of organization, including the interpersonal scale (e.g., there are many ways to transmit the same message between 2 individuals). It can therefore be assumed that this property exists in a collective sports team, considered as a (socio-)biological system made up of interacting players. This means that such a team has many ways to generate the same output or to perform the same function in a given context by using various patterns of interactions between the players. 

Moreover, a system which has many ways to generate the same output in a given context is thus extremely \emph{adaptable} -- that is, the system is able to reorganize its behavior/configuration according to changes in surrounding constraints \citep{edelman2001degeneracy}. In sports context, \citet{Seifert2017} showed that the ability of a system to use various patterns of interactions (while maintaining high efficiency or effectiveness) is a relevant performance indicator. Thus, developing a wider range of (effective) solutions (i.e. exploiting the degeneracy property by using various patterns of interaction between the players) makes the team more likely to be adaptable to changing environmental conditions \citep{araujo2022team}.

Additionally, in complex adaptive systems degeneracy is closely related to complexity \citep{edelman2001degeneracy, whitacre2010degeneracy}. Indeed, \citet{tononi1999measures} showed that degeneracy and complexity are intimately related both empirically (i.e., with a strong positive correlation between information theory measurements of degeneracy and complexity) and conceptually. According to \citet{araujo2022team}, “the more structures within the system that can execute the same functions the more complexity exists within that system”.

Finally, measures used in statistical information theory have been proposed as a framework for evaluating these properties (i.e. degeneracy, complexity) such as entropy and mutual information \citep{tononi1999measures}. It explains the use of entropy measures in the present work: the more various patterns of interaction between the players are used within a team, the more complex is that team (or equivalently: the more that team is degenerated), and the more likely that team is to be adaptable. At the end, these measures are used to characterize and understand the adaptability of the network, as suggested by \citet{tononi1999measures}. Then, the link between the entropy measures and team performance is examined, with this link expected to be positive given that the entropy measure reflects the team’s ability to adapt, which has itself been defined as a relevant performance indicator. Moreover, in information theory entropy measures the uncertainty of a system by calculating the average amount of information in it \citep{Shannon1948}. One can say it describes the size of the space of the states reachable by the system, or how well the system has explored the possible states. In the context of collective sports, there is a tough balance between the organization and disorganization of collective behavior: although the team should maintain organized patterns of behavior to maximize the cooperation between its members, its behavior should also be disordered enough to mislead opponents and maintain enough degrees of freedom \citep{neuman2018adaptive}. Since a game is considered here as a succession of possessions, each being an attempt to score points for the offensive team, the use of various patterns of interaction between players along the possessions can make the offensive team less predictable from the defensive team’s point of view. This constitutes a second reason why the offensive team would benefit from having more complexity in its behavior, and therefore measured by a higher entropy value.

Furthermore, drawing on sociobiological models and the common features of sociobiological systems for team performance analysis makes it possible to explain how repeated interactions between individuals scale to emergent collective behavior \citep{Duarte2012}. Not surprisingly, trying to characterize the playing style of a given team is a classic challenge in collective sports performance analysis. Specifically, the challenge is often to identify regularities in how a team plays or to compare different teams’ organizations (similarities and differences) in order to discuss their respective playing styles -- typically, via network science \citep{Buldu2019}. Studying regularities and comparing team organizations make sense because there is not only one way for a team to respond to a given situation: players can use many coordinative structures, and it is thus interesting to identify how a team tends to act in given circumstances. From an ecological dynamics perspective, the common way in which a team behaves in a given performance context is referred to as the system’s \emph{intrinsic dynamics}. According to \citet{mcgarry2002sport}, it can also be referred to as a \emph{signature}. Thus, the team’s signature can be defined as the preferences/tendencies that emerge as this team is organized when facing a given constraint -- that is, arising from the repetition of interactions between players as they face this specific constraint. In the present work, the relative score between 2 teams (i.e., who is leading and by how many points?) is seen as an evolving constraint shaping the offensive team’s behavior, and therefore its effect on the interaction network is analyzed in order to identify team signatures in the way they adapt to this constraint.

In summary, the purpose of the present work is to model basketball team behavior as a temporal network and quantify its complexity using entropy measures, in order to study the link between the way basketball teams exploit degeneracy in the interactions between players and their collective performance. This is achieved both by investigating the link between the level of complexity in the interaction network’s structure and the final performance of the team and by looking at how the level of complexity in the dynamics of the interaction network structure of each team is affected by its current performance relative to the opponent. These 2 challenges constitute the 2 hypotheses to be tested: (1) the positive correlation between entropy and final score, and (2) the relative score between the 2 teams acts as a constraint that shapes the behavior of the offensive team, possibly in different ways depending on the team. Thus, the remainder of this article is structured as follows. Section \ref{sec2} describes the positioning of this work within the existing scientific literature about network analysis in collective sports and highlights the specific contributions of this work. Section \ref{sec3} provides a general description of the model, an illustration with a single basketball game, and an experimental protocol on a larger dataset in order to test the 2 hypotheses. Last, results of this experimental protocol are presented and discussed in sections \ref{sec4} and \ref{sec5}, and section \ref{sec6} concludes.

\section{Framework}\label{sec2}

\subsection{Related works}\label{sec2sub1}

\subsubsection{Passing networks}
When focusing on interactions between players in a network approach to collective sports, it is usual to look at the successful passes (i.e., when a player gives the ball to a partner): “as a common performance indicator in collective sports, passes appear to be a natural choice to model interactions” \citep{korte2019network}. Such networks are usually called \emph{passing networks} and are often built using all the passes made by one team during one game (or several games). A few authors have studied passing networks in basketball, with different approaches. First, some authors have tried to identify differences between tactical playing positions: they considered each attack as a graph (or equivalently its adjacency matrix) and they summed up these matrices at the game level to compute different network metrics (e.g., in-/out-degree) and tested for differences \citep{korte2018characterizing,clemente2015network}. Second, \citet{Fewell2012a} aimed to quantify and differentiate the offensive strategies of basketball teams, and to do so the authors considered basketball games as transition networks by adding the beginning and the end of each possession to the passing network. Then, they created cumulative networks for each team, and they calculated the different network metrics (e.g., individual and team entropy, individual and team flow centrality) to analyze the various team offensive strategies. Lastly, \citet{xin2017continuous} aimed to identify clusters of players based on their playing style and performance, while revealing differences between teams in terms of offensive strategies. The authors considered basketball games as transactional networks (i.e., instead of observing an edge between 2 nodes, they observed a series of transactions and thus the data simply recorded the senders, the recipients and the time of transactions) and they applied the continuous-time stochastic block model to create these player clusters.

\subsubsection{Temporal analysis of passing networks}
Moreover, in sports in which the goal is to collectively bring the ball to a target (e.g., in the basket), passes can be seen as a flow: thus, each possession can be represented as the trajectory of the ball traveling between players from the recovery to the end of the possession \citep{Fewell2012a,xin2017continuous}. This supports the idea that possession level is relevant \citep{Ramos2018} and therefore that a basketball game can be seen as a succession of possessions. Similarly, \citet{Mattsson2021} proposed a method to extract trajectories in passing networks, treating passes as steps in a walk-process. The authors analyzed these trajectories using summary statistics to replicate the classic findings of sports sciences, and they quantified the complexity of a team’s passing behavior by evaluating the Markov order of that team’s trajectories: they showed that a small but exceptionally successful subset of teams generate complex passing dynamics, which they considered to be evidence of complex multi-player tactics at the top echelon of professional club teams in football \citep{Mattsson2021}. Beyond the analysis of trajectories, which can be used to preserve the temporal dimension of the interactions between the players, other authors have used temporal graphs in sports sciences to take this dimension into account. In football, \citet{yamamoto2011common} built temporal passing networks using 5-minute time windows to analyze network dynamics and identify the properties of the football game as a system. In particular, they studied the probability distribution for the connectivities of the vertices (i.e., players) and the temporal evolution of the hubs (using relative ball-touch frequencies for each player in each time window) to reveal that the soccer game system has features common to other types of complex networks (e.g., self-organization), and they also identified unique features of this system (e.g., the number of triangles formed within the time window reflects the game momentum) \citep{yamamoto2011common}. In handball, \citet{kostakis2017discovering} applied an algorithm they developed on a dataset of a small number of possessions: the algorithm summarizes a temporal network by discovering recurring modes in its dynamics. They modeled a handball team’s activity as a sequence of timestamped edges capturing passes between players, trying to identify a small number of modes, and they segmented time in order to associate each time segment with a mode: by doing so, they were able to reduce the entire set of possessions to a smaller subset of unique possessions \citep{kostakis2017discovering}. Examining the total number of modes that are needed to perfectly summarize a team’s entire gameplay can provide information on the level of diversity in that team’s gameplay, so \citet{kostakis2017discovering} suggested that it would be interesting to study the correlation between this number and team performance in different sports. In rugby, \citet{cintia2016haka} built a multilayer network analysis framework using passes and tackles as the interactions between the players of both teams, and they carried out their analyses on 2 analytical levels: static, by aggregating interactions at the game level, and dynamic, by considering a game as a succession of possessions. When studying multilayer networks at the game level, the authors determined that a rugby team’s passing network should ensure strong connectivity (meaning that there are multiple and reciprocal pathways for the ball to reach all players) and good resilience (as measured by tackle features estimating how much the network is resistant to disruptions) \citep{cintia2016haka}. Then, their analysis of the multilayer networks at the dynamic level partly corroborated these results, but the authors considered that it provided less information than this first analysis on a static aggregated network over the match. This differed from what the previous literature has shown in soccer, and they explained this by the observation that each rugby sequence is part of a grand match strategy (i.e., building on each other to achieve the intended result) and can therefore only be appreciated by analyzing the passing network as a whole (i.e., at game level), whereas soccer sequences yield results that are mostly independent from the other sequences \citep{cintia2016haka}. No studies mobilizing temporal graph theory in basketball have been found but, given that basketball consists of an even faster succession of possessions than football, it is highly likely that looking at the dynamics of the passing network in this sport will provide new information compared to using a static aggregated network.

\subsubsection{Entropy}
Entropy-based measures are widely used in the recent literature in collective sports performance analysis (e.g., \citealp{Gama2020, Martins2020, Martinez2020, Welch2021Collective, Pereira2021}). In all these studies, the authors found that an increase in entropy measures was related to better performance outcomes (as measured by goals scored and/or points earned). In basketball, \citet{Fewell2012a} showed that greater entropy in the collective behavior makes the team less predictable and harder to defend against, which was proved by a positive correlation between the entropy level of a team passing network during a game and the number of points this team scored during this game. To calculate team entropy in the transition network of passes between players, the authors first used Shannon’s entropy at player level to assess the uncertainty of the transitions (i.e., where does the ball go after this player?) and then combined all the individual entropies to calculate the team entropy: they found that team entropy was connected to team success, supporting the hypothesis that a complex and unpredictable ball distribution pattern is an important component in team strategy \citep{Fewell2012a}. However, since their work was based only on pairwise transition probabilities (i.e., when one player has the ball, what is the probability he will pass it to any other player?), they were not really considering actual passing trajectories -- or the temporal dimension in general. In addition, some authors have linked a team’s relative entropy (i.e., measured as the entropy of a team during a game compared to the entropy of the opponent team) and that team’s performance \citep{neuman2018adaptive}. Overall, these findings suggest that entropy-based measures are useful for understanding the coordination dynamics of collective sports teams and can be good predictors of team performance.

\subsubsection{Relative score and team signature}
A part of this work consists in a focus on the effect of the relative score (i.e., who is leading and by how many points?) on the passing network of basketball offensive teams, with the aim of highlighting differences and similarities in the way teams adapt their behavior to this constraint (i.e., their signature). In previous literature, the effect of match status (i.e., winning, losing or drawing) on the length of the passing sequences of top-level football teams was investigated by \citet{paixao2015does}. The authors concluded that teams use longer passing sequences when losing or drawing and shorter when winning, but they also revealed the existence of team signatures of play with regard to this specific constraint, meaning that each team may have its own way to adapt the length of their passing sequences according to match status. In basketball, authors have analyzed the effect of the starting score of the game quarter on the game quarter outcome; see \citet{sampaio2010effects} in top-level men's basketball and \citet{moreno2013effects} in top-level women’s basketball. Even though they did not measure the effect on player/team’s behavior, it is interesting to consider the cut-off values they identified: a split appeared between $8$ and $9$ points for men’s basketball \citep{sampaio2010effects} and between $7$ and $8$ points for women's basketball \citep{moreno2013effects}. These values were then used by the authors to classify quarters between imbalanced (i.e., with greater differences of points between teams) and balanced ones. Also, \citet{zuccolotto2018big} used basketball data to identify high pressure situations by measuring the influence of score difference with respect to the opponent at the moment of the shot (i.e., which is the same variable as the relative score defined in the present work) on the percentage of successful shots. They identified several peaks at $-15, -5, +1, +6$ and $+10$ points \citep{zuccolotto2018big}. These values are useful to note, as they allow comparison with those identified using the model presented in this article.

\subsection{Specific contributions}\label{sec2sub2}
In this work, a temporal passing network model (TPNM) is designed in order to describe collective behavior by looking at interactions between players through the use of temporal graphs. Therefore, the main difference with previous works is the focus on the dynamics of the interactions between basketball players by taking into account the temporal dimension in the passing network, going beyond the analysis of static aggregated networks at the game level. By doing so, it constitutes an attempt to address limitations and respond to suggestions identified in the previous literature.

Indeed, current research using network analysis in collective sports shows some limitations: (1) SNA, as a method to describe team performance, is almost exclusively applied to football, (2) networks are mostly built at game level (i.e., meaning an aggregation of interactions over one entire game), thus limiting the informational value of the analysis, and (3) network approaches hardly consider the dynamics of the interactions \citep{korte2019network}. To address this, the present work focuses on basketball games at the possession level of analysis and includes the temporal dimension of the interactions between the players.

In addition, the suggestions of \citet{Ramos2018} and \citet{Fewell2012a} are taken into account: according to their works, it would be extremely useful to connect networks with temporal models and it would be interesting to explore game dynamics by looking at the effects of constraints (such as the point differential between 2 teams, or the influence of the adversary team) on the network of interactions. Both the design and the use of the TPNM follow these research lines. Indeed, after a first step, which is a successful attempt to replicate a well-known result in sports sciences (i.e., a greater entropy in collective behavior is positively related to performance), the TPNM is used thereafter to identify the little-known effects of the relative score, seen as a constraint shaping the offensive team’s behavior.

Finally, by running this analysis on several teams and comparing the findings, the aim is to identify their respective signatures (i.e., their own way of adapting their behavior to the constraint). Although network analysis has been broadly used in collective sports analysis to characterize and compare team playing styles, \citet{Buldu2019} demonstrated that despite the observation that aggregated passing networks of football teams can show differences between team organizations, the temporal dimension needs to be considered to obtain a more detailed profile of a team in order to identify its signature. It seems that no such study has yet been carried out in basketball. In the present work, the objectives and hypotheses in \citet{Fewell2012a} are retained, but the focus is more on the collective structures of passes rather than on individual players’ pass preferences. The objective here is not to measure the entropy in players’ pass choices, but to use entropy measures to evaluate the use of various patterns of passes by the team, therefore including the temporal dimension but regardless of which individual player is involved in the patterns.

\section{Methodology}\label{sec3}

\subsection{Temporal Passing Network Model}\label{sec3sub1}
In this section, $\mathbb{N}_n$ denotes the set of integers from $0$ to $n$, and $\mathbb{N}^*_n$ denotes the set of integers from $1$ to $n$: $\mathbb{N}^*_n=\{1,2,\dots,n\}$.

\subsubsection{Temporal graph}
A graph is defined as a pair $G = (V, E)$ where $V$ is a set of vertices (or nodes) and $E$ a set of edges (or links) between these vertices. Temporal (or dynamic, according to authors and contexts) graphs have more recently appeared as a way to model dynamic systems and are now the subject of intensive research \citep{Holme2015,Casteigts2018,vernet2023study}. There are several ways to model the dynamics of a network, but the most natural approach is to represent it as a sequence of graphs, where each graph of the sequence -- called a \emph{snapshot} -- represents the relations among vertices at a given discrete time \citep{Casteigts2018}. Here, a rolling time window $T_k$, a snapshot $G_k$, and a temporal graph ${\cal G}$ are defined as follows:

\vspace{1em}
\begin{definition}
\label{Definition 1}
A sequence of time steps $(t_k)_{k \in \mathbb{N}^*_n}$ is defined such that:\\ $\forall k \in \mathbb{N}^*_{n-1}, \ t_{k+1} = t_k+\tau $, where $\tau$ is the step duration and $n$ depends on the length of the given data. At each time step $t_k$, a time window $T_k$ is defined as the time interval $T_k = [t_k, t_k+\delta]$ where $\delta$ is the duration of this time interval. 
\end{definition}

Therefore, this definition of time windows involves 2 configurable parameters: a duration $\delta$ (i.e. the length of the time windows) and a time step duration $\tau$. Normally, we have $\tau < \delta$ which implies overlapping time windows. Moreover, we set $t_1=0$ and $T=t_n + \delta$ the time horizon. Thus, the interval $[0,T]$ is the interval of study.

\vspace{1em}
\begin{definition}\label{Definition 2}
The snapshot $G_k = (V_k, E_k)$ is the graph associated with the time window $T_{k}$, where $V_k$ and $E_k$ are respectively the sets of vertices and edges that exist during $T_k$. 
\end{definition}
In this work, $G_k$ arises as the model of the passing network observed during $T_k$, $E_k$ modeling the set of passes, and $V_k$ the set of players involved. 

\vspace{1em}
\begin{definition}\label{Definition 3}
The temporal graph ${\cal G}$ is defined as the sequence of all snapshots $\{G_1,G_2,...G_n\}$.
\end{definition}
In the context of this work, these snapshots will be classified as graphlets due to their basic and small structure, as explained later.

\subsubsection{Sub-graphs: motifs and graphlets}
Graphs are useful to understand the structure of complex systems by modeling the components and their interactions, but graph theory also supports an understanding of functional features through subgraph (i.e., sub-parts of a graph) analysis. By focusing on small structural patterns of interconnections between the vertices, subgraph analysis constitutes an approach going either (1) beyond a simple pairwise representation of the interactions within a network \citep{Battiston2020} or (2) beyond the global structural features of the network, by focusing on local structural properties in a bottom-up way \citep{prvzulj2004modeling, Milo2002}. Subgraph analysis often consists in the detection of \emph{motifs}, initially defined as the patterns of interactions occurring in complex networks at numbers that are significantly higher than those in randomized networks \citep{Milo2002}. A motif analysis offers insights into the functional dimension of a system: motifs are considered structural signatures of the function of a network, and different motifs can reflect/correspond to different functions or different solutions to the same function \citep{Battiston2020}. Yet a limitation of this approach is that patterns that are functionally important but not statistically significant might exist but would be missed \citep{Milo2002}. Instead of motifs, \emph{graphlets} can be used to address this limitation. In order to avoid any terminology confusion between network motifs and network subgraphs (motifs being special types of subgraphs), authors introduced the term graphlet to define “a connected network with a small number of nodes” \citep{prvzulj2004modeling}. Actually, both motifs and graphlets consist of static, non-directed, and unweighted subgraphs, 2 by 2 non-isomorphic, and classed by the number of nodes. To clarify the difference, one could say that for a given number of nodes p: graphlets constitute a list of all the theoretically possible p-node subgraphs that one may find in any graph (so they are induced subgraphs) when motifs are the subset of significant subgraphs among all the subgraphs identified in a given network (so they are deduced subgraphs). From their basic definitions, motifs and graphlets have both been extended in various ways, for example: 

\begin{enumerate}
    \item motifs on temporal networks, or temporal motifs, modeled as sequences of temporal edges within given time windows \citep{Paranjape2017} or as a succession of snapshots \citep{Oberoi2023}
    \item motifs on sequential/trajectory data, or sequential motifs, via observed walks on directed and sequence-ordered graphs \citep{LaRock2022}
    \item motifs in weighted networks, or weighted motifs, detected by random walk \citep{Picciolo2022}
    \item graphlets on temporal networks, either static-temporal graphlets (for a snapshot-based approach) or dynamic graphlets \citep{Hulovatyy2015}
    \item graphlets on directed networks, or directed graphlets, adding the edge direction and so significantly increasing the number of graphlets \citep{Aparicio2017}.
\end{enumerate}

The field of subgraph analysis is very active and often involves an extension of the original concepts (i.e., motifs, graphlets) to particular data (i.e., with specific features). In this work, the choice was to list all possible graphlets, a priori, with a given number of nodes and respecting the specific data restrictions. As this work involves temporal graphs with timestamped and directed interactions, more possible graphlets than the original classification (i.e., static, non-directed, and unweighted) exist, but with the application of specific data restrictions there may actually be fewer possibilities than the directed graphlet classification.

\subsubsection{Network profiles}
Many works on subgraphs share the same objective -- that is, to count subgraph occurrences (i.e., subgraph frequency) in order to describe the structure of networks and discriminate/compare them. As it is possible to talk about a motif profile to characterize the collection of statistically validated frequencies with which the various motifs are observed in a network \citep{Battiston2020}, the collection of frequencies of each graphlet identified will be referred to as a \emph{graphlet profile}. the collection of the frequencies of each graphlet identified will be referred to as a graphlet profile. If one possession is enough to create a profile, one can also create a profile by aggregating possessions according to a given criterion (e.g., profile of a team). Work has already been done in this direction with sports data, using motifs of passes in football in order to create a passing network profile, which was then used to describe and/or compare the playing style of teams and/or individuals \citep{ 
bekkers2019flow, malqui2019soccer, meza2017flow}. However, these works are fundamentally different from the present one since they were based on passing trajectories (as described earlier), whereas here the graphlet profile build is the result of a snapshot-based approach. The aim of such an approach is to better fit the dynamics of players’ interactions, but some authors have mentioned a limitation: a certain lack of information remains because there is no consideration of the inter-snapshot relationships \citep{Hulovatyy2015}. For this reason, a \emph{transition profile} is created to complete the graphlet profile. The transition profile is defined as the matrix representing all the inter-graphlet transition frequencies between 2 consecutive snapshots. As well as the graphlet profile, the transition profile is a collection of frequencies that can be created by aggregating possessions: it should be noted that in such cases there would be no artificial link created between the last state of one possession and the first state of the next one since a transition matrix is created for each possession and the aggregation consists in an addition of matrices. From a complexity science perspective, this can be seen as a states and transitions problem: a graphlet in a given time window corresponds to one state of the system, and a transition is the succession of 2 states (for 2 consecutive time windows). It is formalized as follows: 

\vspace{1em}
\begin{definition}\label{Definition 4}
The state space of the system corresponds to the set of all occurring graphlets. Let $N$ be the number of graphlets that can be reached by the system. The state space is denoted by ${\cal E}  = \{g_1, g_2 ... g_N\}$, where $g_i$ denotes graphlet $i$, with $i \in \mathbb{N}^*_N$. A prior probability $p_i$ is associated with each state $g_i$, computed from the observed frequency of the occurrence of $g_i$. The prior graphlet distribution is revealed by the vector of probabilities $(p_i)_{i \in \mathbb{N}^*_N}$.
\end{definition}

\begin{definition}\label{Definition 5}
Let $X_k$ be the random variable associated with the state of the system in the $T_k$ window. On the whole time interval $[0,T]$, the evolution of the system can be modeled as the Markov chain $(X_k)_{k \in \mathbb{N}^*_n}$ governed by the following transition probabilities: 
\[
\forall k \in \mathbb{N}^*_n, \forall (i,j) \in (\mathbb{N}^*_N)^2, \ p_{ij} = P(X_{k+1} = g_j | X_k = g_i)
\] 
Each value $p_{ij}$ is computed from the observed frequency of transition from $g_i$ to $g_j$ between all successive time windows. The Markov chain is “learned” by analyzing a set of possessions (the whole game, the whole competition, \dots). The transition profile is then identified with the matrix of transitions of the chain: $M= (p_{ij})_{1\leq i,j \leq N}$. 
\\The sum of the values on each line of this matrix equals $1$; that is \\
$\forall i \in \mathbb{N}^*_N, \ \sum_{j=1}^N \ p_{ij}=1$.
\end{definition}

One could also consider a transition only when there is a change of state in the system (i.e., when 2 consecutive time windows are associated with different states of the system, or equivalently when $i\neq j$ in the transition matrix). This leads to a third profile called the \emph{restricted transition profile}, created in the same way as the transition profile. Here again, the restricted transition profile is a collection of frequencies, and it can be created by aggregating possessions (in the same way as the transition profile):

\vspace{1em}
\begin{definition}\label{Definition 6}
Let us consider the probability of transition $p'_{ij}$ between $g_i$ and $g_j$ when only transitions with a change of state are considered, leading to a restricted transition matrix $M'= (p'_{ij})_{1\leq i,j \leq N}$.\\
This matrix therefore undergoes some modifications:\\
$\forall i \in \mathbb{N}^*_N,\ p'_{ii}=0$, and $\forall i\neq j,\  p'_{ij}= p_{ij}/(1-p_{ii})$ so as to get $\forall i \in \mathbb{N}^*_N, \ \sum_{j=1}^N \ p'_{ij}=1$.\\
The restricted transition profile is identified with the matrix $M'$.
\end{definition}

The combination of the 3 profiles (i.e. graphlet profile: see Figure \ref{Figure_3}, transition profile and restricted transition profile: see Figure \ref{Figure_4}) is expected to capture the dynamics of the interactions between players more accurately than a single graphlet profile would, thus providing a more comprehensive description of the behavior of the system. 

\subsubsection{Entropy measures for the TPNM}
To measure the level of complexity in the behavior of such a system, entropy-based metrics are used (as introduced previously). Each metric is associated with one profile, and it is worth noting that their definition can be applied to any system modeled by a Markov chain.

\vspace{1em}
\begin{definition}\label{Definition 7}
Let $SE$ denote the State Entropy, $TE$ the Transition Entropy, and $RTE$ the Restricted Transition Entropy. These entropies are defined as follows:
		\begin{equation*}
		\begin{array}{ll}
			SE & = -\sum\limits_{i=1}^N p_i \log_2(p_i)\\
		 TE & = -\sum\limits_{i=1}^N p_i \sum\limits_{j=1}^N p_{ij} \log_2(p_{ij}) \\
		RTE & = -\sum\limits_{i=1}^N p_i \sum\limits_{j=1,j\neq i}^N p'_{ij} \log_2(p'_{ij})
		\end{array} 
		\end{equation*}
where $p_i$ is the prior probability of state $i$, $N$ is the total number of reachable states, $p_{ij}$ is the probability of transition from state $i$ to state $j$, and $p'_{ij}$ is the transition probability from state $i$ to state $j$ excluding self-transitions; these quantities are defined in definitions \ref{Definition 5} and \ref{Definition 6}.
\end{definition}

Although it seems that these metrics have not been used in previous literature to study a succession of interaction patterns between players, they have been used for instance to analyze eye movement by modeling transitions between areas of interest as a state-transition problem \citep{krejtz2015gaze}. Authors have used both “stationary entropy” and “transition entropy” to describe the behavior. In the present work, the same stationary entropy is used, but 2 transition entropies are differentiated depending on whether or not what they call a “self-transition” (i.e. $i = j$) is taken into account. Thus, “state entropy” (or SE), “transition entropy” (or TE), and “restricted transition entropy” (or RTE) are used to refer to these 3 metrics (Definition \ref{Definition 7}). According to the interpretation of these metrics by \citep{krejtz2015gaze}, it can be assumed here that: (1) a higher value of SE means that the team distributes its behavior more equally among graphlets and a lower value is obtained when behavior tends to be concentrated on certain graphlets, and (2) the higher TE or RTE is, the more randomness there is in a team’s transitions, the more complex the sequence graphlets are, and the more “surprise” there is in the collective behavior. As justified in the previous sections, it is assumed here that a greater entropy measures a greater level of complexity in the behavior of the offensive team, therefore reflecting a greater exploitation of their degeneracy property, which is expected to be linked to better performance outcomes.

\subsection{Illustration of the methodology on 1 basketball match}\label{sec3sub2}
In the previous subsection, a model and 3 associated metrics were defined in a theoretical perspective, and the current subsection shows an application on a restricted real dataset (i.e., 1 basketball game). The next subsection will detail the experimental protocol set up to test the 2 hypotheses -- introduced in section \ref{sec1} -- that are tested on the complete dataset. Results will then be reported and discussed in the following sections. 

\begin{figure} 
    \centering
    \includegraphics[width=1\textwidth,keepaspectratio]{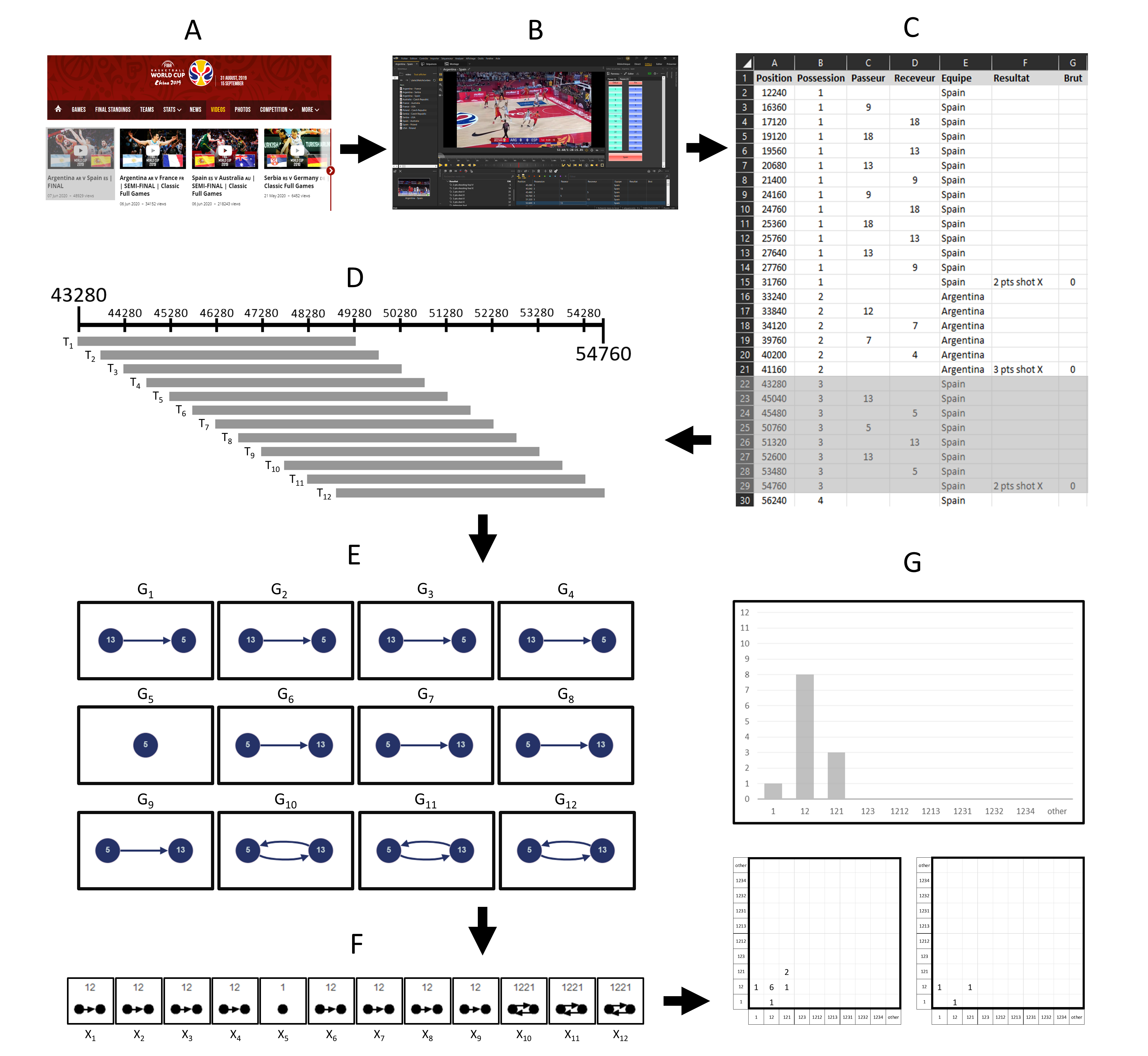}
    \caption{Step-by-step data workflow. (A) Get the videos from the FIBA website. (B) Annotate the videos with game events and contextual information. (C) Compile all data in a csv file. (D) Define all discrete time windows (definition \ref{Definition 1}) for each possession. (E) Create the snapshot for each time window (definition \ref{Definition 2}) and thus the temporal graph (definition \ref{Definition 3}) of each possession. (F) Associate each snapshot with a state of the system (definition \ref{Definition 5}). (G) Generate a graphlet profile, transition profile and restricted transition profile by counting their respective occurrences in each possession. Possessions can then be accumulated according to different criteria (e.g. team) to create profiles associated with these criteria. Dartfish software is used up to step C, then R programming is used through the RStudio environment.}
    \label{Figure_1}
\end{figure}

\begin{figure}[H] 
    \centering
    \includegraphics[width=0.75\textwidth,keepaspectratio]{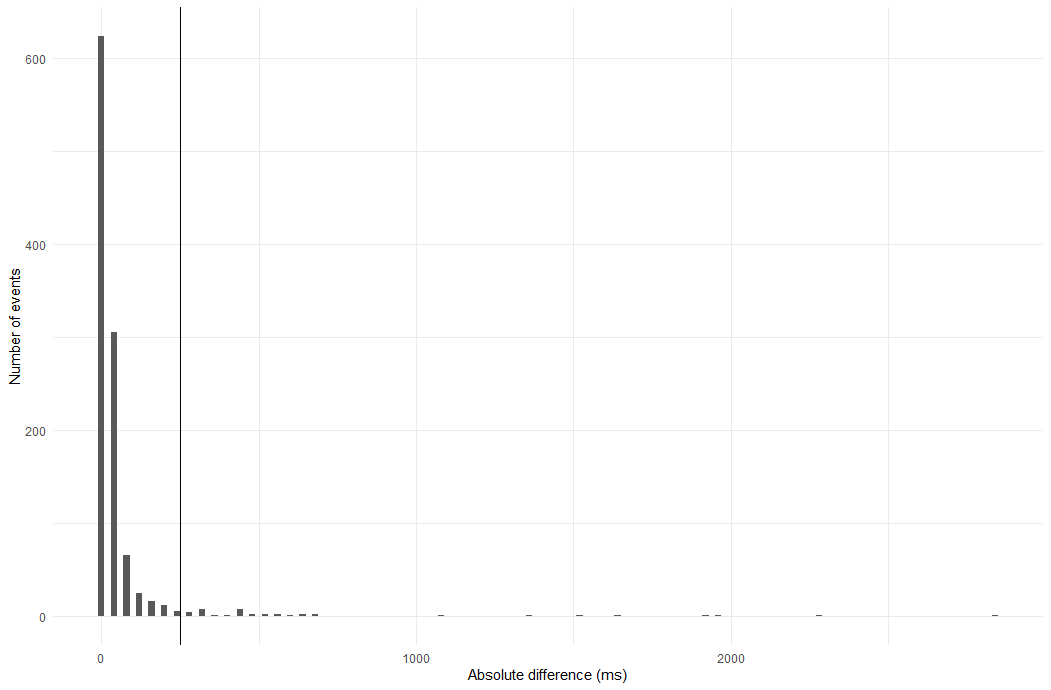}
    \caption{Absolute difference between timecodes for reliability analysis. $10\%$ of the events (i.e., passes and receptions) in the dataset were randomly selected and manually recorded a second time, in order to compare these timecodes with the timecodes of the same events in the original data. The figure shows the distribution of absolute differences for all 1092 events that were compared in pairs. Each bar corresponds to 40 milliseconds of error, which matches the frame rate of the videos. The black vertical bar is displayed as a reference at $0.25$ seconds, which corresponds to the step of the time windows, $96.4\%$ of events are annotated with less than $0.25$ seconds absolute difference.}
    \label{Figure_13}
\end{figure}

\begin{figure}[H] 
    \centering
    \includegraphics[width=0.75\textwidth,keepaspectratio]{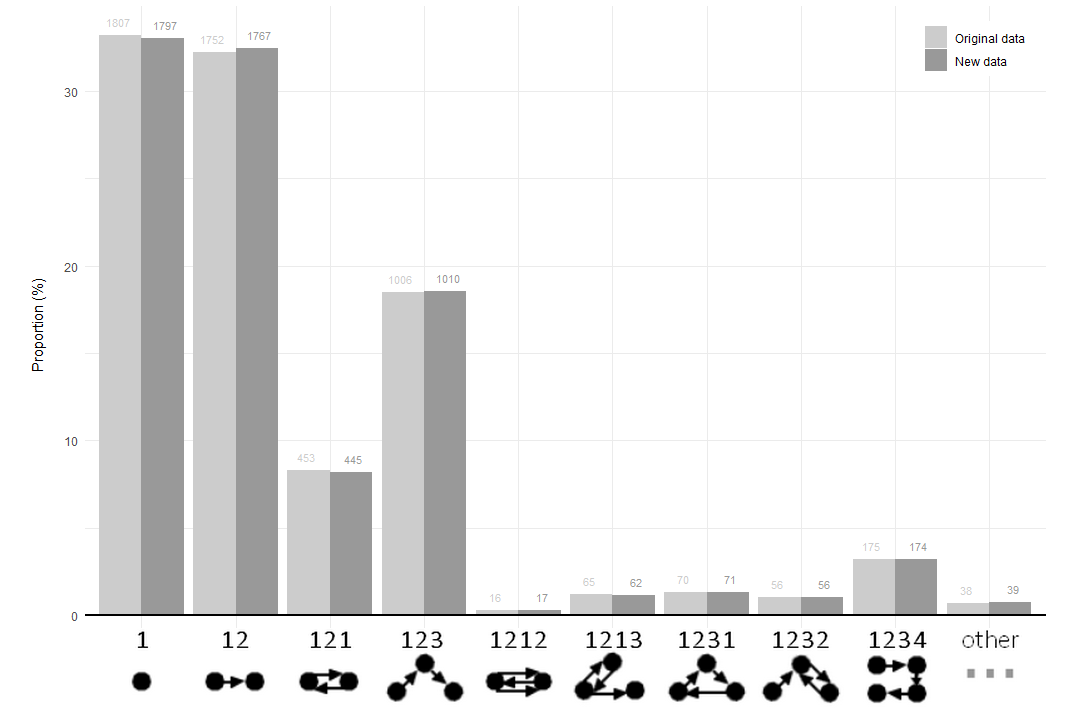}
    \caption{Graphlet distribution comparison for reliability analysis. "New data" corresponds to the randomly selected subset representing $10\%$ of the dataset that has been manually recorded a second time, and "Original data" represents the same $10\%$ of the dataset but from the original dataset. The figure shows the number of graphlets resulting when using a time window with the following parameters: $\delta = 6$ seconds (duration) and $\tau = 0.25$ seconds (step). The chi-square test of independence indicates no difference in proportion ($\chi^2$ = 0.295, N = 10876, df = 9, p = 1.000).}
    \label{Figure_14}
\end{figure}

In this work, the model is applied to a dataset of 12 games from the 2019 men’s FIBA Basketball World Cup (i.e. all games from the quarter finals). This involved 8 teams, each playing 3 games. Using videos available on FIBA’s YouTube channel (Figure \ref{Figure_1}A) and using the Dartfish software, each successful pass (and its timecode) was recorded manually (Figure \ref{Figure_1}B) by one observer. A reliability analysis has been performed by repeating the manual recording on a randomly selected subset representing $10\%$ of the dataset: it revealed that event timecodes differ by less than 0.25 seconds in $96.4\%$ of cases and that the error percentage is less than $1\%$ in graphlets profiles (Appendix, Figure \ref{Figure_13} and Figure \ref{Figure_14}). Additional information required for the analysis was also recorded: the start and end timecodes of each possession (to be able to reconstruct the entire timeline), the relative score at the start of the possession (which is considered a constraint shaping the offensive team's behavior), and the outcome of the possession (to assess the functional performance of the system). Raw data represents 6096 passes divided into 2213 possessions within the 12 games (Figure \ref{Figure_1}C). This subsection focuses on the final game of the competition (i.e., Argentina vs Spain) to illustrate the methodology. 

The rolling time window is applied to each possession (Figure \ref{Figure_1}D), with a $\delta = 6$ seconds duration and a step of $\tau = 0.25$ seconds. These parameters are based on the nature of the data: the time scale of the dynamics of the network should not be too far from the time scale of the dynamic system \citep{Holme2015} and it has been found that $6$ seconds is the best fit with the interaction dynamics in a basketball offensive team: a smaller duration prevents the occurrence of sufficiently complex graphlets, and a larger duration results in too few time windows. In addition, possessions that last less than the window duration are automatically removed from the analysis (i.e., the time window duration is like a filter removing the shortest possessions) and a duration of $6$ seconds is commonly used to discriminate fastbreaks (i.e., short possessions in which the offensive team tries to take advantage of a defensive imbalance by playing fast) from other possessions: previous research has shown that $9$ of $10$ fastbreaks last less than $6$ seconds \citep{Cardenas2015} or that fastbreak duration is mostly between $3$ and $6$ seconds \citep{Courel-Ibanez2017}. Thus, a $6$-second duration also seems appropriate because the present work focuses on longer possessions (as this is where rich patterns of interaction between players can emerge). This 6-second value is probably specific to modern basketball as played according to FIBA rules, in particular given the current 24-second shot clock (i.e. with a few exceptions, a team has a maximum of 24 seconds to shoot from the moment it gets possession of the ball). It is interesting to note that, historically, there have been regulatory changes concerning this shot clock duration that speed up the pace of the game, making possessions shorter and more numerous within a game \citep{vstrumbelj2013decade}. In other leagues in which this rule is different (e.g. 30 seconds in NCAA) or if there is any modification of this rule, the 6-second value would probably no longer be appropriate. The $0.25$-second step has been chosen to fit data granularity: the shortest time between 2 passes is around $0.28$ seconds, which means that using a step of $0.25$ seconds ensures that no more than 1 interaction is ever added or removed between 2 consecutive time windows, thus allowing an effective assessment of interaction dynamics.

\begin{figure}[H]
    \centering
    \includegraphics[width=0.9\textwidth,keepaspectratio]{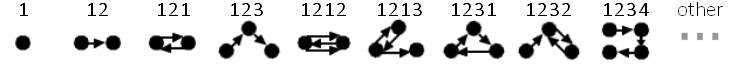}
    \caption{The list ${\cal E}$ of states reachable by the system (definition \ref{Definition 4}) with their label: “1” is the graphlet with 0 edge, “12” to “1234” are the 8 possible graphlets with 1 to 3 edges, and “other” contains all graphlets with at least 4 edges.}
    \label{Figure_2}
\end{figure}

The nature of the dataset and the chosen parameters limit the number of possible graphlets and transitions. First, there are only $8$ graphlets with 1 to 3 edges, which are the same as the $8$ motifs generated with a walker allowed to move for a maximum of $3$ steps \citep{Picciolo2022}. Indeed, the transmission of an object (e.g., a ball) between individuals constitutes a directed and ordered interaction, as is a walk or a trajectory. In line with the literature, the graphlets issued from the data used in this work could have been called sequential graphlets, but for the sake of language simplicity they will be referred to as graphlets throughout this paper. Also, as there is an exponential increase in the counting complexity when the number of maximum steps increases, there is a need to fix a maximum number of steps and to find a compromise between this complexity and the motif size \citep{Picciolo2022}. Here, the decision was made to consider the $8$ graphlets with $1$ to $3$ edges and to add the graphlet with no edge (i.e., reflecting a $6$-second time window in which a player kept the ball).The particularity of the data is that there are very few interactions between very few players, but with a fast pace. Since within a given time window there can only be a limited number of interactions, it is possible to consider each snapshot as a graphlet (Figure \ref{Figure_1}E, Figure \ref{Figure_1}F). Therefore, a set compiling all remaining possible graphlets (i.e., all graphlets with $4$ edges or more) is also considered, meaning that $n = 10$ states is considered to be reachable by the system, labeled ${\cal E} = \{ 1, 12, 121, 123, 1212, 1213, 1231, 1232, 1234, other \} $ (see Figure \ref{Figure_2}).

\begin{figure}[H]
    \centering
    \includegraphics[width=0.8\textwidth,keepaspectratio]{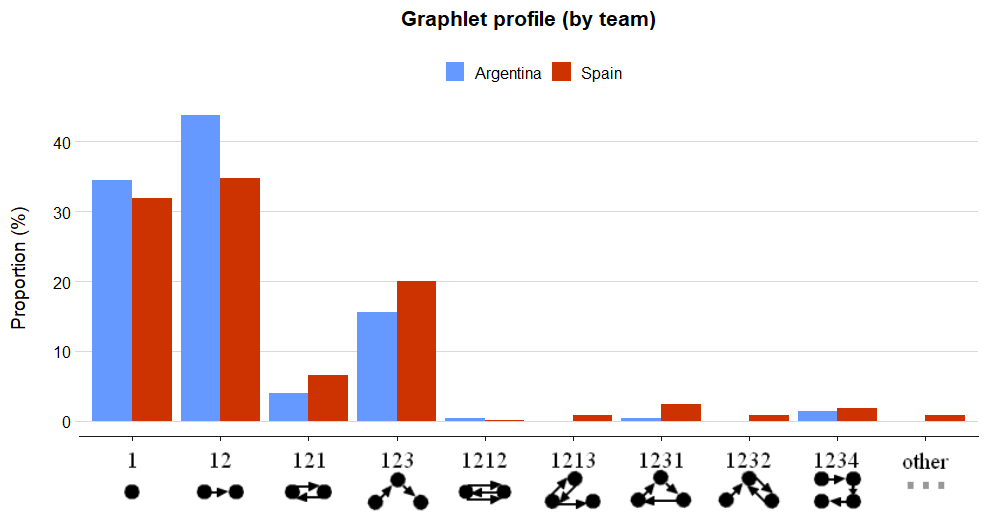}
    \caption{Graphlet profile by team during the final. Values correspond to $p_i$ (i.e., probability of state $i$) but in percentage.}
    \label{Figure_3}
\end{figure}

\begin{figure}[H]
    \centering
    \includegraphics[width=0.7\textwidth,keepaspectratio]{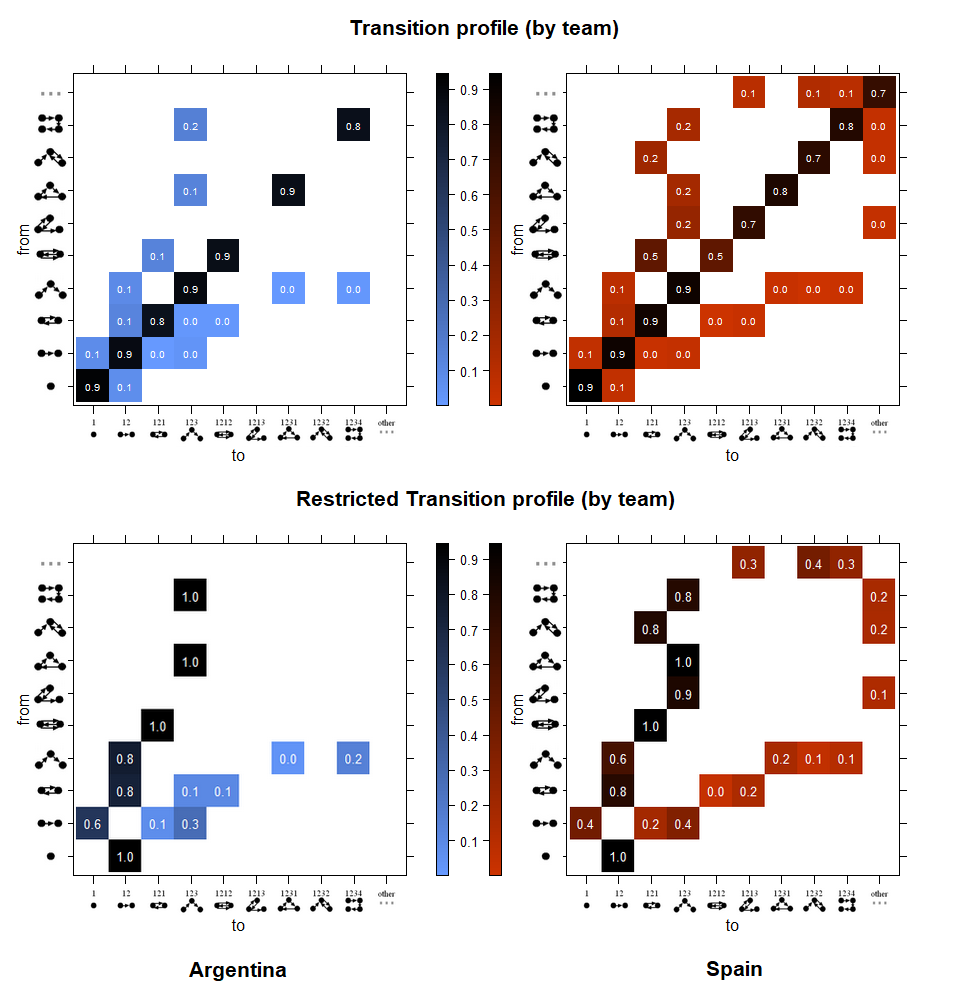}
    \caption{Transition profile (at the top) and restricted transition profile (at the bottom) by team (Argentina in blue on the left, Spain in red on the right) during the final. Each transition is to be read from a graphlet on a row to a graphlet on a column, and for each transition $p_{ij}$ is displayed (i.e., probability of the second graphlet at $t+1$, knowing the first graphlet is at $t$).}
    \label{Figure_4}
\end{figure}

\begin{figure}[H]
    \centering
    \includegraphics[width=0.7\textwidth,keepaspectratio]{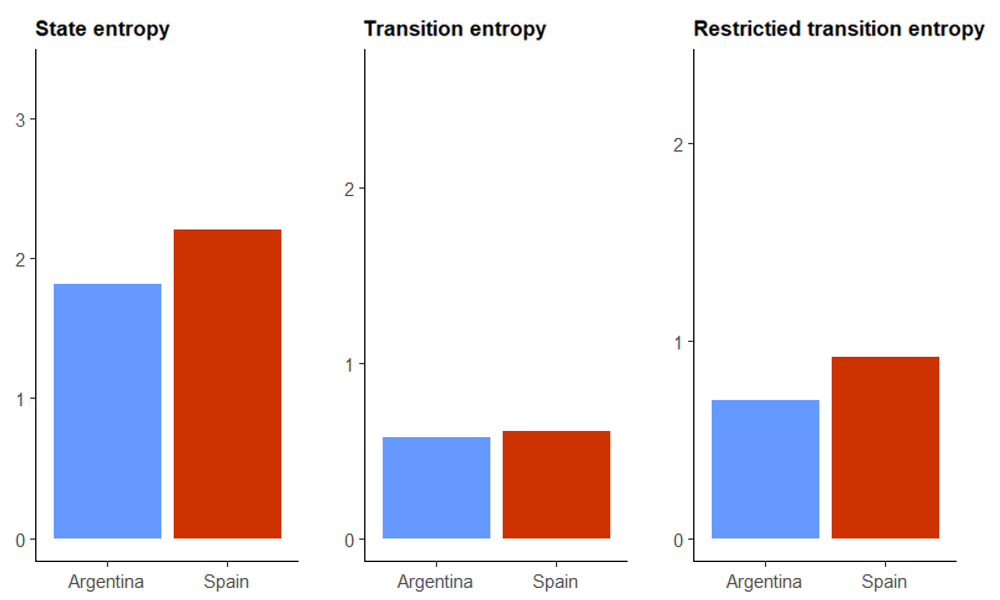}
    \caption{Entropy levels on the 3 metrics (from left to right: state entropy, transition entropy, restricted transition entropy) for both teams (Argentina in blue, Spain in red) during the final. The y-axis goes from the minimal entropy (that is 0 for the 3 metrics) to the theoretical maximal entropy.}
    \label{Figure_5}
\end{figure}

At this point, each possession is considered as a succession of states. The next step is then to build profiles for states, and for transitions (Figure \ref{Figure_1}G). Figure \ref{Figure_3} shows the graphlet profile of both teams during the final: this means that all graphlets over all possessions of each team have been aggregated to build both profiles. In the same way as for graphlets, there are certain practical limitations on the transitions that are actually possible. Indeed, because the step between 2 consecutive time windows that has been chosen is smaller than the time between 2 passes, all the theoretical transitions (and so restricted transitions) are not possible (e.g. if $i$ is $121$, then $j$ cannot be $1231$). Figure \ref{Figure_4} shows transition and restricted transition profiles, built in the same way as the graphlet profile (i.e. aggregating and counting). Under such conditions, maximal entropy values are around 3.322, 2.658, and 2.356 for state entropy, transition entropy, and restricted transition entropy, respectively. Considering SE, this maximum entropy is reached if all state probabilities are equal. For TE and RTE, this maximum entropy is reached if, for a given state, the transition to each of the other possible states has the same probability, and that this is the case for each state. Figure \ref{Figure_5} illustrates the entropy level of both teams in the 3 metrics calculated for the final.

During the final (in which the Spanish team won $95-75$ against the Argentinian team), the Argentinian team used graphlets “1” and “12” proportionally more (and “1212” to a lesser extent) than the Spanish team, which used proportionally more of all the other graphlets -- the more complex ones (i.e., involving more players and/or with more passes). In addition, the Spanish team’s graphlet profile has a greater entropy, meaning that this team used more various patterns of interaction than the Argentinian team. The same result appears for both transition and restricted transition profiles: the matrices exhibit more various transitions for the Spanish team than for the Argentinian team. The transition entropy metric does not show a very clear difference between the 2 teams, but a higher level of entropy for the Spanish team can be identified by comparing the restricted transition profiles of the 2 teams. This can be explained easily by the parameters of the model: a small step ($\tau = 0.25$ second) that allows only 1 pass more or 1 pass less was chosen, so it is very likely that 2 consecutive snapshots will consist of the same graphlet (i.e., $j = i$). Thus, the weight of the diagonal overwrites any possible differences, and removing it by excluding self-transition (i.e., $j \neq i$) revealed some actual differences. However, the best-performing team was the one with more entropy in its behavior and such results are in line with the literature. More particularly, the results indicate that the Spanish team used both more various patterns of interaction and more variation in the sequence of patterns than the Argentinian team, and this link with performance may be explained by the fact that doing so offers greater potential for adaptability and greater unpredictability from the point of view of the other team. This hypothesis is tested in the next section, on the entire dataset.

Moreover, it can be observed that the 2 least complex graphlets (i.e., “1” and “12”) together account for more than $50\%$ of all graphlets, and this for both teams. Nevertheless, it does not mean that more complex graphlets are negligible: it would not be surprising if, although rare, such patterns were crucial and had a considerable impact on the game dynamics.

\subsection{Experimental protocol}\label{sec3sub3}
The purpose of the experimental protocol is to test the 2 hypotheses presented in the introduction, on the entire dataset. As the data are not normally distributed, non-parametric statistical tests are used. Also, the level of significance is set at $p = .05$.

The first part consists in analyzing the relationship between the network of interactions (and here more especially its level of complexity measured by entropy metrics) and its functional outcome (i.e., the result of the possession). This approach is a robust way to perform network analysis, according to \citet{Fewell2012a}. For testing the first hypothesis, Spearman’s rank correlation coefficient (i.e., non-parametric measure of association between 2 variables, \citealp{mccrum2008correct}) is calculated between points scored in a game and entropy metrics. In addition, the Wilcoxon signed-rank test (i.e., non-parametric test used for paired samples, \citealp{mccrum2008correct}) tests whether or not an entropy greater than that of the direct opponent is related to winning.

The second part consists in analyzing the effect of the relative score on the dynamics of the interaction network, in order to understand how the system adapts to variations in this constraint. This fits precisely with the ecological dynamics framework according to \citet{Correia2013}. Here, possessions are split into classes according to the relative score at the start of each possession. First, in a supervised approach, entropy levels are compared between 5 pre-determined classes corresponding to: a big deficit, a small deficit, a balanced situation, a small advantage, and a large advantage for the team in possession of the ball. The number of points per possession (i.e., ratio) is computed for each class: this is used as a performance indicator. The chi-square ($\chi^2$) test (i.e., test used to compare proportions between 2 groups, \citealp{mccrum2008correct}) is used to test the differences between the graphlet profile of each class. Second, in an unsupervised approach, possessions were split into 3 not pre-determined classes based on the maximization of entropy differences between the classes (Appendix, Algorithm \ref{Algorithm_1} is a description of the algorithm). Here again, the points per possession ratio is used to assess performance for each class, and the same $\chi^2$ test is used to compare classes according to their graphlet profiles. But this time, the level of entropy per class is calculated for each team separately: this way it is expected to identify each team’s signature (i.e., to precisely capture the effect of the constraint on each team and therefore the way in which each team adapts according to changes in this constraint). Also, the differences between the teams’ classes in terms of both entropy and performance are tested using the Wilcoxon signed-rank test.

In summary, the dataset is analyzed to assess the relationship between entropy and performance: first, by investigating the link between a greater entropy and success in the game, and second, by analyzing the effect of the current performance as a dynamical constraint on a team’s behavior and its entropy level (with a focus on differences and similarities between teams).

\section{Results}\label{sec4}

\subsection{Entropy and final performance}\label{sec4sub1}
To analyze the relationship between the level of entropy in a team's passing behavior and the team's performance, the game is used as the scale of analysis. Thus, there is a total of 24 observations (i.e., 12 games $\times$ 2 teams per game). For each observation, 3 profiles (i.e., graphlet, transition, restricted transition) are built by aggregating all the possessions played by the team during the game. Then, entropy is measured by calculating the 3 metrics (i.e. state entropy, transition entropy, restricted transition entropy). Performance is evaluated here as the number of points scored by the team during the game (i.e., their final score). First, the correlation between the entropy variables and the performance variable is tested. Second, a comparison is made between each team's entropy level for each pair of 2 direct opponents, therefore comparing the entropy of the winning team with that of the losing team.

    \subsubsection{Correlation with points scored by the team}\label{sec4sub1.2}
\begin{figure}[H]
            \centering
            \includegraphics[width=0.9\textwidth,keepaspectratio]{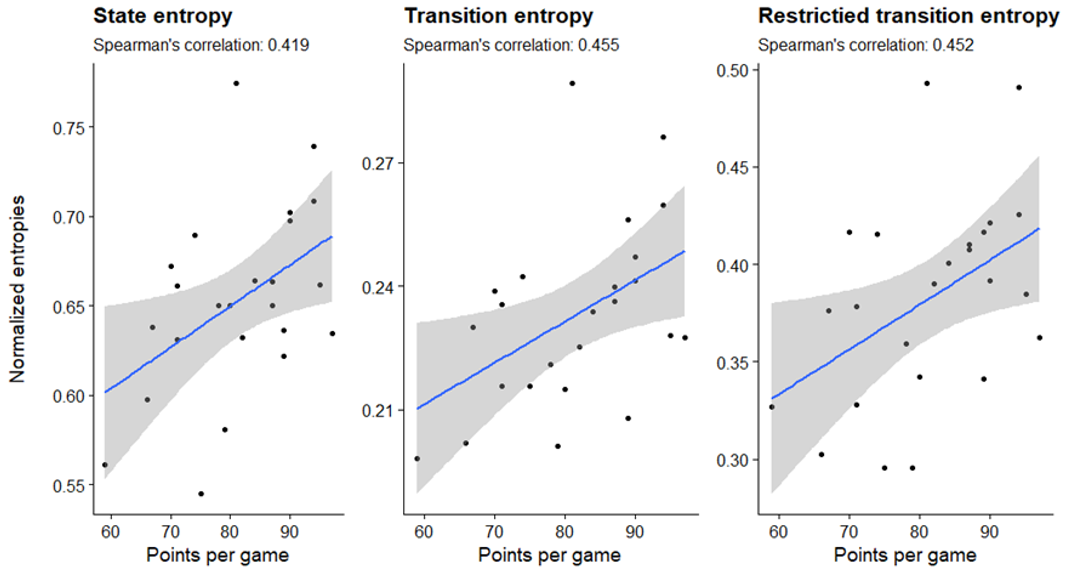}
            \caption{Correlation plots between the performance variable (points per game) and the 3 entropy variables (SE, TE, RTE) normalized by their respective maximum. Each dot is 1 game by 1 team, the linear regression between the 2 variables is displayed in blue and the 95\% confidence interval in gray.}
            \label{Figure_6}
        \end{figure}

    There is a significant correlation (at $p = .05$) between the performance variable and each of the 3 entropy metrics, with p-values ranged from $0.026$ to $0.042$ (Table \ref{Table_1}). Spearman's rho indicates that these correlations between the performance variable and each of the 3 entropy metrics are positive and moderate, with p-values ranging from $0.42$ to $0.46$ (Table \ref{Table_1}). Thus, at the game level, there is a moderate significant positive correlation between the level of entropy calculated on the temporal passing network of the attacking team (for SE, TE and RTE) and the number of points scored by this team (Figure \ref{Figure_6}).

        \begin{table}[H]
            \caption{Spearman's rank correlation coefficient and p-value between the performance variable (points per game) and each one of the 3 entropy variables (from left to right: SE, TE, RTE).}\label{Table_1}
            \begin{tabular}{@{}llll@{}}
            \toprule
             & State entropy & Transition entropy & Restricted Transition entropy \\
            \midrule
            Spearman’s rho      & 0.419	& 0.455	& 0.452 \\
            p-value             & 0.042* & 0.026* & 0.027* \\
            \bottomrule
            \end{tabular}
            \raggedbottom{* significant at p $\leq .05$}
        \end{table}

    \subsubsection{Relative comparison between winner and loser}\label{sec4sub1.3}
\begin{figure}[H]
            \centering
            \includegraphics[width=0.9\textwidth,keepaspectratio]{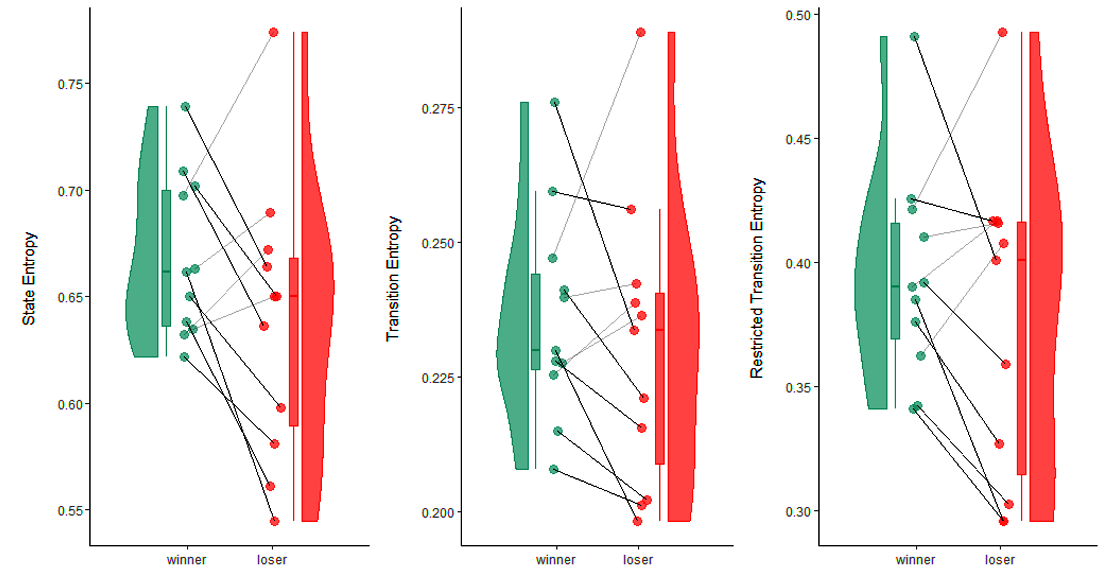}
            \caption{Raincloud plots for the 3 entropy variables (from left to right: SE, TE, RTE) comparing the winner (in green) and the loser (in red) of each game. Line between the winner and the loser is black when it is the winner who has the greater entropy, and gray when it is the loser.}
            \label{Figure_7}
        \end{figure}

    The relative difference between the winner and loser of each game (Figure \ref{Figure_7}) shows that the winners had greater entropy in their collective behavior 63.6\% of the time (7 of the 11 games, 1 game was excluded because it ended with a tie). Meanwhile, the Wilcoxon signed-rank test with “Winner $>$ Loser” as the alternative hypothesis shows no significant differences ($p=0.062$ for SE, $p=0.183$ for TE, $p=0.139$ for RTE).
    
\subsection{Relative score and entropy}\label{sec4sub2}
\begin{figure}[H] 
        \centering
        \includegraphics[width=0.9\textwidth,keepaspectratio]{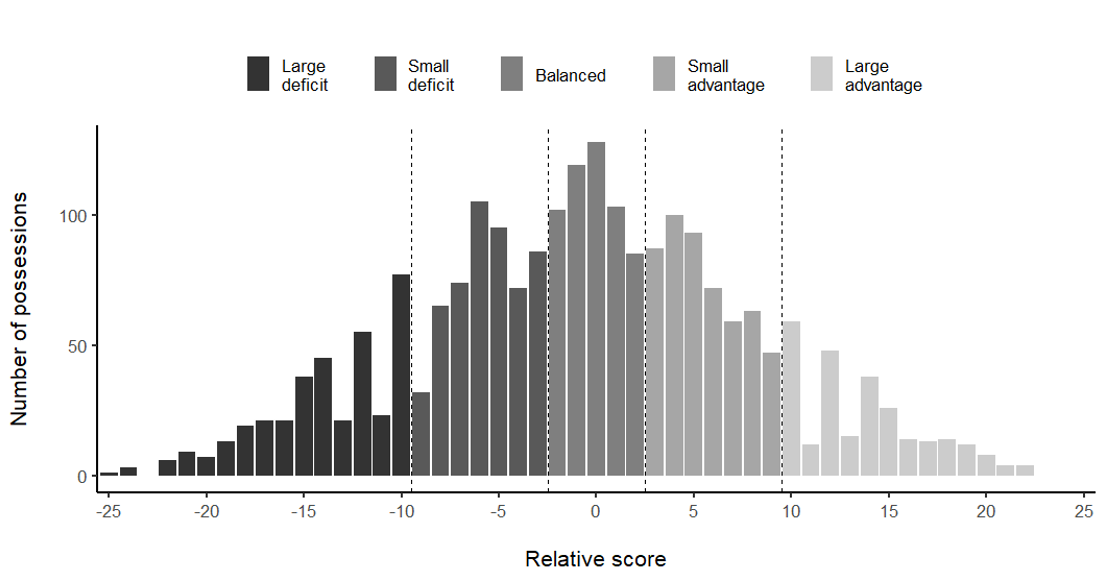}
        \caption{Distribution of possessions by their relative scores (i.e., score of the offensive team minus score of the defensive team, at the beginning of the possession) with the 5 classes displayed.}
        \label{Figure_8}
    \end{figure}

To analyze the relationship between the current performance and the level of entropy in each team’s passing behavior, the possession is used as scale of analysis. A “relative score” value is attributed for each one of the 2213 possessions played over the 12 games. This discrete variable is defined from the point of view of the attacking team, thus calculated as the score of the offensive team minus the score of the defensive team at the beginning of the possession. Then, possessions are classified into 5 classes according to their relative score value, corresponding to: a big deficit, a small deficit, a balanced situation, a small advantage, and a large advantage (Figure \ref{Figure_8}, Table \ref{Table_2}).

    \begin{table}[H]
        \caption{Description of the 5 classes of relative score  by their boundaries (1st line) and by the number and the proportion of possessions within them (2nd line).}\label{Table_2}
        \begin{tabular}{@{}llllll@{}}
        \toprule
         & Large deficit & Small deficit & Balanced & Small advantage & Large advantage \\
        \midrule
        Relative score      & –25 to –10 & –9 to –3   & –2 to +2   & +3 to +9   & +10 to +22 \\
        Number (proportion) & 359 (16\%) & 529 (24\%) & 537 (24\%) & 521 (24\%) & 267 (12\%) \\
        \bottomrule
        \end{tabular}
    \end{table}
    
When considering only the possessions that last for 6 seconds or more (i.e., excluding fastbreaks, free throws, etc.), 1751 possessions remain (79\% of all possessions) and the proportions of possessions per class do not change, as the chi-square test of independence indicates no difference in proportion ($\chi^2$ = 0.348, N = 3964, df = 4, p = 0.987).
    
    \subsubsection{Entropy and performance metrics comparison by class}\label{sec4sub2.1}
    These 5 classes are used to design 5 profiles in order to analyze each team's passing behavior according to the relative score. Chi-squared independence tests indicate significant differences (at $p = .05$) between graphlet profiles of all classes except 2 (i.e., between small deficit and large advantage), with p-values less than or equal to $0.004$ (Table \ref{Table_3}). Then, the 3 entropy variables (i.e., SE, TE, RTE) are calculated for each class (Table \ref{Table_4}) and the number of points per possession is used as the performance indicator, calculated as the number of points divided by the number of possessions within a given class. Results indicate that SE, TE and RTE are all maximized in the small advantage condition (i.e. relative scores from $+3$ to $+9$) whereas the performance variable (i.e. points per possession, for each class) is greater in the large deficit condition (i.e. relative scores from $-25$ to $-10$).
    
        \begin{table}[H]
        \caption{Chi-squared independence tests between the 5 classes of relative score, 2 by 2.}\label{Table_3}
            \begin{tabular}{@{}lllll@{}}
            \toprule
             & $\chi^2$ & N & df & p-value \\
            \midrule
             Large deficit - Small deficit & 26.166 & 20239 & 9 & .002* \\
             Large deficit - Balanced & 37.051 & 20405 & 9 & \textless .001* \\
             Large deficit - Small advantage & 28.394 & 20127 & 9 & \textless .001* \\
             Large deficit - Large advantage & 24.530 & 14489 & 9 &.004* \\
             Small deficit - Balanced & 46.588 & 26622 & 9 & \textless .001* \\
             Small deficit - Small advantage & 66.315 & 26344 & 9 & \textless .001* \\
             Small deficit - Large advantage & 16.474 & 20706 & 9 & .058 \\
             Balanced - Small advantage & 32.143 & 26510 & 9 & \textless .001* \\
             Balanced - Large advantage & 40.203 & 20872	& 9 & \textless .001* \\
             Small advantage - Large advantage & 45.375 & 20594 & 9 & \textless .001* \\
            \bottomrule
            \end{tabular}
            \raggedbottom {* significant at p $\leq .05 $}
        \end{table}
    
        \begin{table}[H]
        \caption{The 3 entropy variables (from left to right: State entropy, Transition entropy, Restricted Transition entropy) and the performance variable (number of points per possession) calculated by class.}\label{Table_4}
            \begin{tabular}{@{}lllll@{}}
            \toprule
             & SE & TE & RTE & Pts/poss \\
            \midrule
             Large deficit & 0.657 & 0.239 & 0.333 & \textbf{0.949} \\
             Small deficit & 0.659 & 0.234 & 0.355 & 0.876 \\
             Balanced & 0.667 & 0.237 & 0.357 & 0.769 \\
             Small advantage & \textbf{0.676} & \textbf{0.240} & \textbf{0.371} & 0.868 \\
             Large advantage & 0.669 & 0.230 & 0.356 & 0.875 \\
            \bottomrule
            \end{tabular}
        \end{table}
        
    \subsubsection{Unsupervised approach to class possessions, at team levels}\label{sec4sub2.2}
    Although the use of 5 pre-determined classes already shows differences, an unsupervised approach constitutes a more detailed analysis to better study the effect of relative score on a given team's passing behavior. Thus, a preliminary step was to identify relevant classes, for each team individually. To do so, all possible classifications (i.e., combinations of classes) with 3 classes of relative score have been computed, with a single limiting rule: each class must represent at least 10\% of the data (Appendix, Algorithm \ref{Algorithm_1}). Then, the 3 entropy variables have been calculated for each of the 3 classes, and this for each classification. Finally, the “maximum $-$ minimum difference” between the 3 classes of each given classification (i.e. the entropy value of the class with the highest entropy minus the entropy value of the class with the smallest entropy) was calculated for each entropy variable. This metric was used to sort the classifications, considering that the best one was the one maximizing this maximum $-$ minimum difference. This process has been repeated for each team separately. The performance indicator is the same as the one previously used (i.e., points per possession).
    
\begin{figure}[H]
            \centering
            \includegraphics[width=1\textwidth,keepaspectratio]{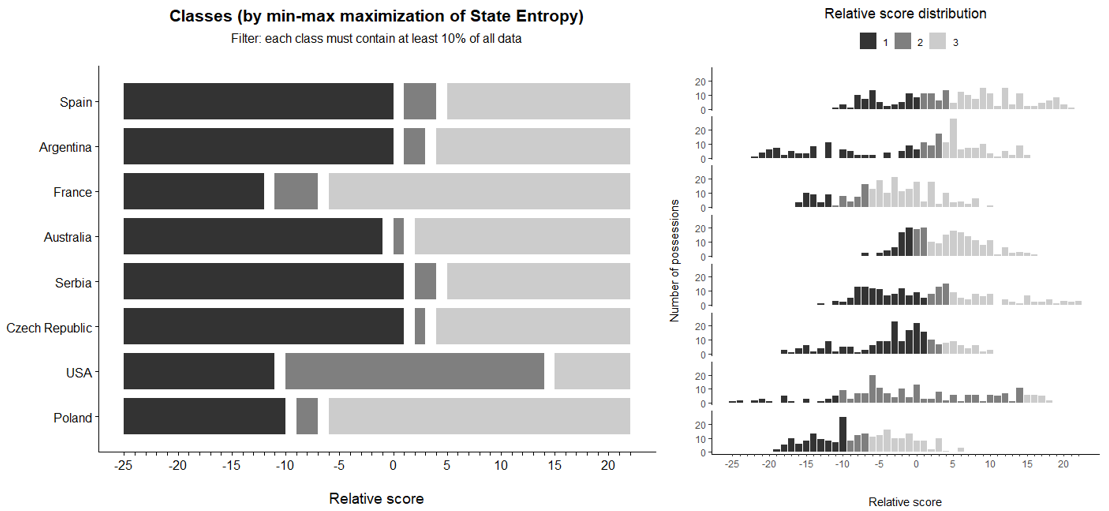}
            \caption{The classification of possessions into the 3 classes that maximize the “maximum $-$ minimum difference” (i.e., the entropy value of the class with the higher entropy minus the entropy value of the class with the smallest entropy) in state entropy. Left part: the 3 classes of relative score, by team. Right part: distribution of possessions within the 3 classes, by team.}
            \label{Figure_9}
        \end{figure}

\begin{figure}[H]
    \centering
    \includegraphics[width=1\textwidth,keepaspectratio]{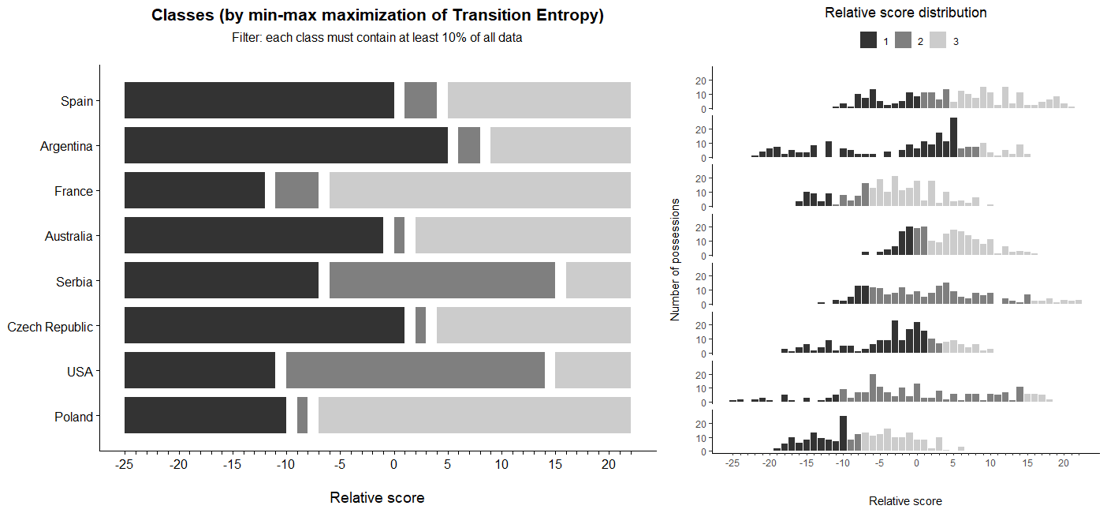}
    \caption{Similar to Fig. \ref{Figure_9} but for transitions: the classification of possessions into the 3 classes that maximize the ”maximum – minimum difference” in Transition entropy.}
    \label{Figure_10}
\end{figure}

\begin{figure}[H]
    \centering
    \includegraphics[width=1\textwidth,keepaspectratio]{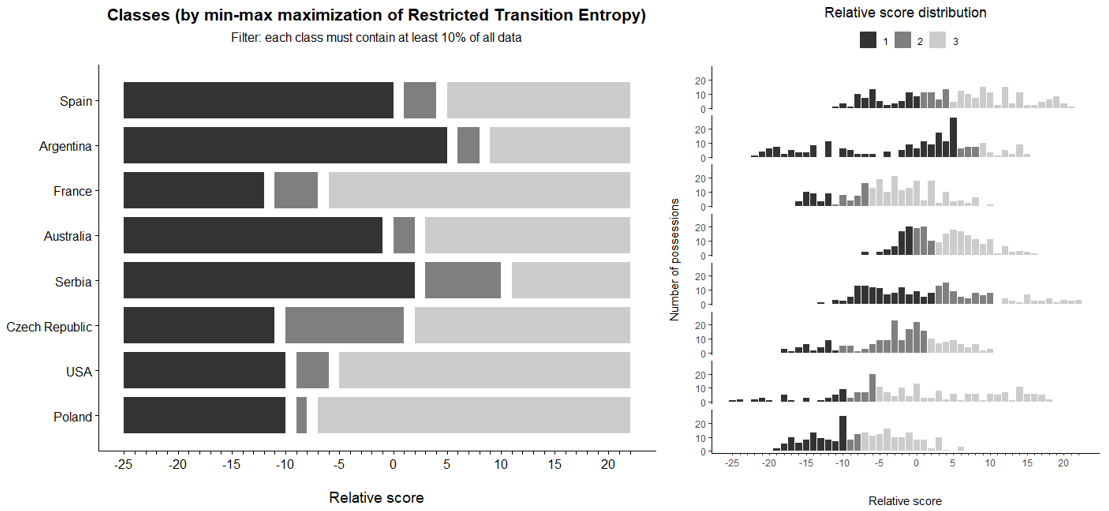}
    \caption{Similar to Fig. \ref{Figure_9} but for restricted transitions: the classification of possessions into the 3 classes that maximize the ”maximum – minimum difference” in Restricted transition entropy.}
    \label{Figure_11}
\end{figure}

    Figure \ref{Figure_9} shows the best classification of each team to maximize SE, as well as the possessions that each class represents (Appendix, Figure \ref{Figure_10} for TE; Appendix, Figure \ref{Figure_11} for RTE). In the same way, Table \ref{Table_5} shows the entropy level and Table \ref{Table_6} the points per possession calculated for these classes. Note that in Figure \ref{Figure_9} and in both Table \ref{Table_5} and Table \ref{Table_6} the teams are sorted according to their final ranking: the first at the top and the last at the bottom.
    
    The Wilcoxon signed-rank test is used to check whether there are significant differences between the 3 respectively ordered classes of each team (i.e. tier respective lower, middle and upper classes) in terms of state entropy and in terms of points per possession (Table \ref{Table_7}). Results of the test indicate that there is no significant difference for the points per possession (at $p=0.05$), but that the lower class has a statistically lower entropy than the middle class and upper classes, with p-values of $0.023$.
    
        \begin{table}[H]
        \caption{State entropy calculated for each class of the classification maximizing the “maximum $-$ minimum difference” (i.e., the entropy value of the class with the higher entropy minus the entropy value of the class with the smallest entropy) of each team.}\label{Table_5}
            \begin{tabular}{@{}llll@{}}
            \toprule
             Team & lower class & middle class & upper class \\
            \midrule
          Spain & 0.612 & \textbf{0.731} & 0.665 \\
      Argentina & 0.547 & \textbf{0.667} & 0.645 \\
         France & 0.528 & \textbf{0.685} & 0.621 \\
      Australia & 0.615 & \textbf{0.668} & 0.610 \\
         Serbia & 0.666 & \textbf{0.759} & 0.695 \\
 Czech Republic & 0.702 & \textbf{0.818} & 0.779 \\
            USA & 0.675 & 0.616 & \textbf{0.706} \\
         Poland & 0.652 & \textbf{0.746} & 0.653 \\
            \bottomrule
            \end{tabular}
        \end{table}
        
        \begin{table}[H] 
        \caption{Points per possession for each class of the classification maximizing the “maximum $-$ minimum difference”  of each team.}\label{Table_6}
            \begin{tabular}{@{}llll@{}}
            \toprule
             Team & lower class & middle class & upper class \\
            \midrule
            Spain          & 0.830 & \textbf{1.020} & 0.908 \\
            Argentina      & 0.829 & 0.917 & \textbf{0.991} \\
            France         & 0.778 & \textbf{0.955} & 0.788 \\
            Australia      & \textbf{0.896} & 0.523 & 0.811 \\
            Serbia         & 0.973 & \textbf{1.090} & 0.879 \\
            Czech Republic & 0.855 & 0.667 & \textbf{1.020} \\
            USA            & 0.917 & \textbf{0.959} & 0.591 \\
            Poland         & 0.809 & 0.879 & \textbf{0.927} \\
            \bottomrule
            \end{tabular}
        \end{table}
        
\begin{table}[H] 
    \caption{Test de Wilcoxon}\label{Table_7}
    \begin{tabular}{@{}lllllll@{}}
        \toprule
        & \multicolumn{3}{l}{Entropie de l'état} & \multicolumn{3}{l}{Points par possession} \\
        \cmidrule(lr){2-4} \cmidrule(lr){5-7}
        & W & z & p & W & z & p \\
        \midrule
        Classe inférieure \& classe moyenne & 2.000 & -2.240 & 0.023*  & 14.000 & -0.560 & 0.641 \\
        Classe inférieure \& classe supérieure & 2.000 & -2.240 & 0.023*  & 15.000 & -0.420 & 0.742 \\               
        Classe moyenne \& classe supérieure & 29.000 & 1.540 & 0.148   & 20.000 & 0.289 & 0.844 \\
        \cmidrule(lr){1-7}
    \end{tabular}
    \raggedbottom {* significatif à p $\leq$ .05}
\end{table}

\section{Discussion}\label{sec5}

\subsection{Linking structure and function: relationship between entropy and final score}\label{sec5sub1}
According to the first hypothesis, entropy in a team’s behavior is positively correlated with the final performance. This correlation is significant, but at a moderate level: it is still an important result because it is not simply anecdotal to find a correlation, even moderate, between a behavioral variable and a variable that is directly related to team performance. However, this result fits with the literature in collective sports performance analysis. More particularly, as was the case for the 3 entropy measures, it means that the diversity of states (i.e., interaction patterns) and the diversity of transitions between states (including when only the transitions with an effective change of state are considered) are linked to a higher number of points scored in the game. Thus, basketball teams that distribute their behavior more equally between graphlets and whose transitions have more randomness are more likely to score more points over the course of the game.

This link between entropy and performance may be explained by 2 mechanisms: a team, as a system, using more various patterns of interaction is (1) more adaptable to changes in the surrounding constraints, which is a requirement for performance, and (2) less predictable from the point of view of the defensive team, which is also a good thing from a performance perspective. This is possible thanks to the degeneracy property of complex adaptive systems, and, as expected, a team exploiting the degeneracy property by using various patterns of interaction between the players (i.e., exhibiting a higher level of complexity in its behavior) is more likely to be more successful. Although no author has explicitly established this link between a basketball team’s usage of the degeneracy property and its ability to score points, some of the literature might be relatable to this. Indeed, by defining basketball as a network problem where each possession is a pathway to score points, \citet{skinner2010price} conceptually demonstrated that basketball teams might benefit by using various pathways rather than focusing only on the most successful ones. To be more precise, considering one possession at a time and using the most successful pathway is a short-term sub-optimal vision (which would be analogous to the Nash equilibrium strategy in game theory) because sacrificing the immediate success of a given play by using another pathway can lead to an increase of global team efficiency \citep{skinner2010price}. Thus, one could extend this statement by saying that the optimal strategy for a basketball offensive team requires exploiting the degeneracy property and that this would be reflected by a higher level of entropy.

\begin{figure}[H] 
    \centering
    \includegraphics[width=1\textwidth,keepaspectratio]{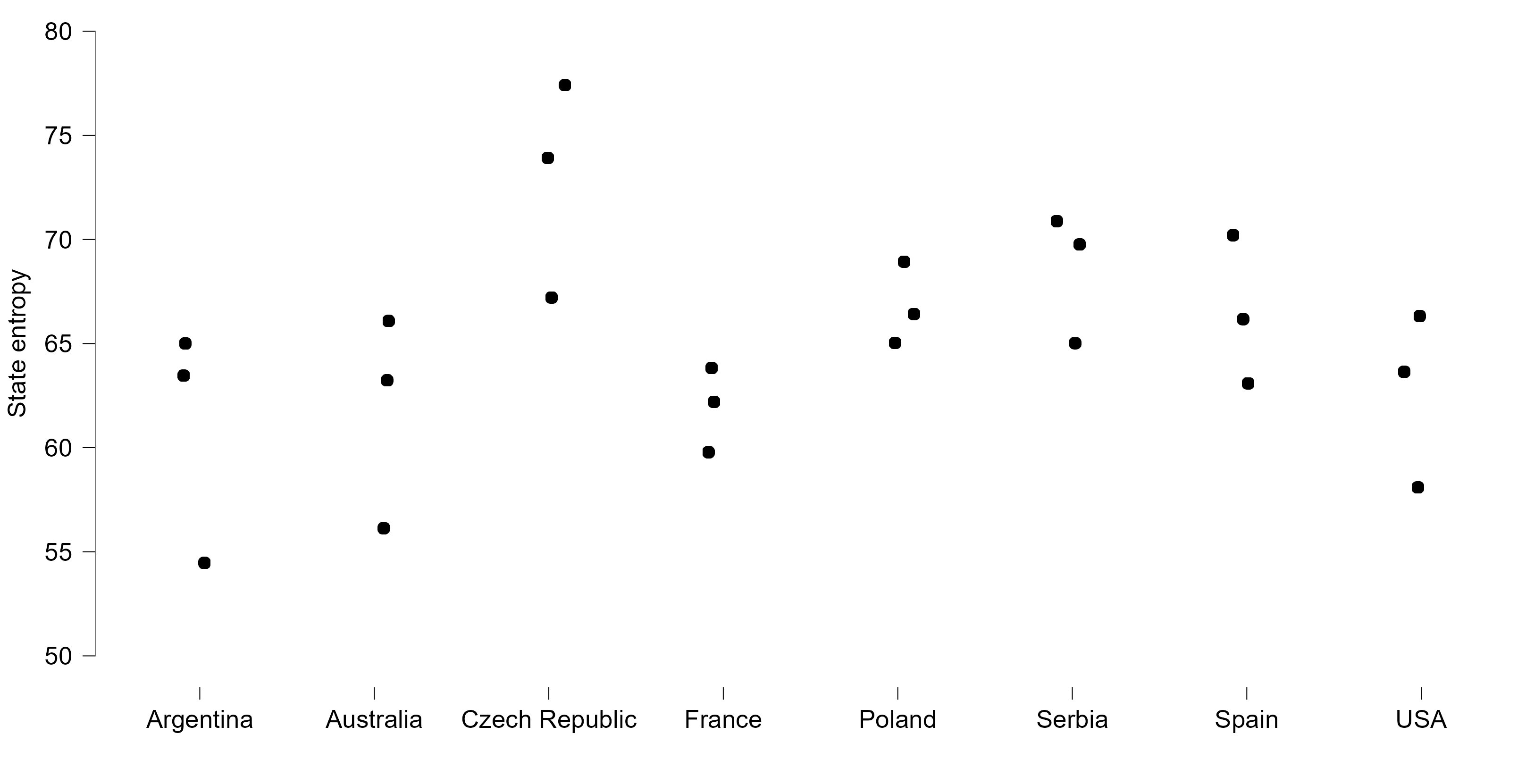}
    \caption{State entropy for the 3 games of each team (in percentage of the maximum).}
    \label{Figure_12}
\end{figure}

However, the other result found in the literature that was tested (i.e., that greater entropy relative to the direct opponent is linked to performance) seemed more difficult to corroborate. There is no significant difference between the entropy in a winner’s passing behavior and the entropy in the loser’s passing behavior, but it would be interesting to test this hypothesis on a larger dataset since there is a tendency in this direction. Particularly, there is 1 game that seemed notably different, going completely in the opposite direction to the hypothesis. This is the game between Serbia and Czech Republic, in which Serbia won, but in which Czech Republic had $77.41$ in normalized state entropy entropy compared to $69.76$ for Serbia. In fact, it turns out that Czech Republic tended to have a higher level of entropy than the other teams: when the average normalized state entropy of a team over a game was $65.26$ (with a standard deviation of $5.22$) all 3 games of the Czech Republic team revealed a higher value: $77.41$, $73.91$, and $67.21$ (Appendix, Figure \ref{Figure_12}). Thus, it seems that this particular team played with a different style, more based on the use of various patterns of passes between players compared to other teams, therefore exhibiting a greater level of entropy. Following the hypothesis of this work, this playing style should have been more likely to increase team performance, but the Czech Republic team lost 2 of their 3 matches and only finished 6th in the competition. This can be interpreted in conjunction with the literature: when \citet{Fewell2012a} used network metrics to capture potential offensive strategies, one of which was the strategy to distribute the ball as a way to reduce predictability, this was captured with a high “team entropy” (i.e., entropy of the transition matrix describing ball movement probabilities across the 5 players and the 2 outcome options). Their results indicated that team entropy was quite related to team performance, but the authors further concluded that, since each network measure could capture a different dimension of team strategy, it would be more interesting to combine several of them (in particular entropy, centrality and clustering) in order to understand team strategy \citep{Fewell2012a}. This is to say that the present work focuses on the relationship between team performance and entropy, but the fact that entropy may -- or may not -- be a component of a team’s strategy has been overlooked. 

While this work relies on the ability of a complex system to be degenerate, the understanding of the link between system structures and functions remains limited. Indeed, a key corollary of degeneracy is that the same structure can, in different circumstances, yield different functions: this is known as \emph{pluripotentiality} \citep{mason2015hidden}. Where degeneracy implies a many-to-one structure-function mapping, pluripotentiality implies a one-to-many structure-function mapping \citep{friston2003degeneracy}. In this work, it has been assumed that a basketball offensive team has only one function, which is to score points, whereas at the possession level other functions may exist (e.g., a team leading when the match is almost over may want to stall for time). Therefore, for a better understanding of the structure-function mapping, which is inextricably linked to system’s complexity, it may be important to consider not only various structures but also various functions of the system. It should also be noted that this is not the only way to approach the analysis of a collective sports game from a complex systems perspective: one could also consider all the players on the field as parts of a single system, within which there would be cooperative interactions (i.e., between team members) and competitive interactions (i.e., between opponents).

\subsection{Effects of a constraint: relationship between relative score and entropy}\label{sec5sub2}
With regard to the second hypothesis, considering the current score -- or more precisely the current point differential with the opposing team -- as a constraint led to interesting insights into the offensive team’s behavior, including how the relative score can affect the level of entropy in a team’s behavior or the identification of critical values of the relative score as common and unique team signatures.

With possessions grouped into 5 pre-defined classes of relative score, differences in the proportion of graphlets used can be observed. As a result, these differences appear at the entropy level: the maximal level of entropy appeared when the offensive team had a small advantage, indicating that more various passing behaviors appeared when the offensive team played with $+3$ to $+9$ points more than the defensive team. However, this does not appear to be linked with performance since the offensive teams scored more points in a large deficit situation. Yet this result is easily understandable because a team with a large lead on the scoreboard often takes advantage of this situation to rest its best players, which makes it easier for the opposing team. This has actually been identified in the literature: \citet{sampaio2010effects} found that when a team had a big lead at the beginning of a quarter time, the opposing team was more likely to recover points during the following quarter. However, there were only minor differences and the interests of such an approach remain limited.

To better understand how teams adapt their passing behavior to the relative score constraint, it appears to be more relevant to classify possessions in an unsupervised way and with a focus on each team separately. Indeed, as each team may well have its own way of adapting to this constraint (i.e., have its own signature) and, what is more, as all teams do not face the same constraints in the same proportions (i.e., some lead more often, others are led more often), the appropriate classifications of possession by relative score are necessarily different. Examination of these classifications can then help to answer several questions: Is it possible to identify teams sharing similar classifications? Are there common values of relative score which appear to be common boundaries between the classes? Do some teams have a unique signature? 

Actually, 3 types of classifications can be identified: the first one was shared by 5 teams (namely: Spain, Argentina, Australia, Serbia and Czech Republic), the second by 2 teams (France and Poland), and a third only for one team (USA). By analyzing team signatures, it is thus possible to identify teams adapting their behavior in a similar way, and teams with their own unique way of adapting their collective behavior to the relative score constraint. But beyond these teams’ intrinsic dynamics, it may also be very interesting to look at the critical values of the constraint (i.e., the boundaries between the classes) to better understand how these values affect team behavior. These values were computed for each team, but shared ones can be identified. For example, the relative score value that most commonly constituted a boundary between classes, shared by $5$ teams out of the $8$, is at $+1$ point. Thus, this relative score value can be considered as a threshold that creates a change in the collective behavior of most teams. A second threshold value, shared by $4$ teams, can be identified at $+4$ points. Other values would be around $-11$/$-10$ (shared by 3 teams), $-7$/$-6$ (shared by 2 teams) and at $+14$/$+15$ (for only 1 team). Although \citet{zuccolotto2018big} did not study the effect of relative score on passing behavior but on shooting performance, it is still instructive to compare the critical values they found to those that have been identified here: they found thresholds values at $+1$ (which is the main threshold value identified here) and at $-5$ (close to one of those identified here), and also possible ones at $-10$ (as identified here) and at $+6$ (not far from one of those identified here). 

Once the most relevant classifications specific to each are determined, it is also possible to compare teams according to their respective first, second and third classes (or lower, middle and upper classes) in terms of entropy and performance. It is worth noting that for all teams except 1 (namely: USA), the middle class is the one with the highest entropy level of the 3 classes. There is probably a link with the observation that this team also has a very different classification from the others. Further, a significantly lower entropy is observed in the lower class than in the middle and upper classes, meaning that teams tend to play with less variety in their interaction pattern in their respective lower class of relative score. Finally, each class can be associated with a performance level by calculating the points per possession within each class. In this case, the class in which a team scores the most points per possession is either in their middle class (for 4 teams) or in their top class (for 3 teams), but rarely in their bottom class (only 1 team), which is a class with significantly less entropy. Despite this tendency, no clear and direct link can be established between performance and entropy in each team’s respective classes. Parameters other than entropy may help to explain the performance level of a class: as stated earlier, a team with a large positive or negative relative score may play in a different way (e.g., with less motivation, with other players, \dots).

\subsection{Temporal Passing Network Model: benefits and limitations}\label{sec5sub3}
The aim of this work was to describe the behavior of basketball teams by designing a Temporal Passing Network Model (TPNM) that integrates the temporal dimension into the analysis of players’ passing network. Thus, it starts with a reduction of a basketball team’s behavior to a temporal graph (first building a succession of snapshots, then converting them into a succession of graphlets) on which entropy measures are calculated (more precisely, on the collection of graphlets and the collection of transitions between graphlets). This is done to evaluate the level of complexity in the system behavior, which reflects the extent to which the system degeneracy property was exploited. Therefore, it only provides partial knowledge about the overall collective behavior. Meanwhile, the model allowed to test 2 hypotheses and led to results in line with those reported in the literature (i.e. a positive relation between the level of entropy in collective behavior and team performance, critical relative score values identical or close to those identified by previous authors). While this does not constitute a formal validation of the model, it is a positive indication of its good relevance. This work introduced a new model in collective sports performance analysis, but further work is needed to validate it more thoroughly.

Nevertheless, there are multiple interesting ways to improve the model. Notably, it should be possible to add more information to the graph (e.g., move beyond the pass and integrate other kinds of interactions, include the spatial dimension, consider players’ roles, take the opposing team into account, \dots) and the analysis could also be improved (e.g., analyze temporal motifs -- which can be defined in this case as a given succession of graphlets -- or moving beyond the degeneracy property by using metrics other than entropy). Other limitations of this work can be noted here. The main one is probably the small number of data items (i.e., 12 games, 2213 possessions, 6096 passes). Although this was sufficient to develop the model and run a first analysis, the findings cannot be generalized to all basketball games: they might be specific to the European international teams that played the final bracket of the 2019 World Cup. For example, previous authors have observed that complex patterns in passing interaction can appear in clubs but not so much in international football teams \citep{Mattsson2021}. Thus, the results might be different in other competitions -- in particular, the more complex graphlets might appear more frequently -- so more data would be required, including other types of population (e.g., women’s basketball, NBA teams, young elite players). Also, the use of more data -- and in particular in longitudinal designs -- would provide deeper insights into team signatures since it should be possible to identify constant features in team behaviors across games or championships in matches with a variety of teams. Moreover, the method presented in this work is only observational, whereas it might be very useful to set up an interventional protocol with the goal of manipulating the constraints whose effects one wants to study.

\section{Conclusion}\label{sec6}

By analyzing the attacking team as a complex system and using possession as the scale of analysis, it was possible to add context to the data, making it easier to study the effect of a given constraint on the emerging collective behavior or on the link to the outcome: this emphasizes the benefits of the ecological-dynamics framework. Moreover, the Temporal Passing Network Model (TPNM) presented in this work offers a way to analyze collective passing behavior while integrating temporal information, therefore being more in line with the actual dynamics of the system’s behavior. This model constitutes a new representation of the collective sports team, based on graphlet and transition profiles. The profiles can be used to identify the team's signature, and this model proves to be effective in identifying differences between teams. While further work is needed to test actual validity of the model, this provides an indirect validation.

By modeling team behavior as a temporal graph, the purpose of this work was to investigate the following 2 questions: Is the amount of entropy in a team’s temporal network of interactions linked to the final performance of that team? And does the current performance act as a constraint affecting the exploitation of degeneracy (i.e., level of entropy) in players’ interactions? In line with the first hypothesis, a positive correlation was identified between the amount of entropy measured on the temporal graph used to model the passing behavior of the team during a game and the number of points scored by that team during that game. This is possibly because playing with more varied patterns of interaction offers greater potential for adaptability -- which is allowed by the inherent degeneracy feature of complex systems -- and because it generates more unpredictability from the point of view of the other team. Second, a link between the relative score and changes in the complexity level of a team’s passing behavior was characterized. In short, the combination of all the results indicates that a team tended to have the greatest level of entropy in its behavior in situations of small scoring advantage and tended to have the lowest level of entropy in those situations making up the fraction of its worst possessions (from the perspective of its score relative to the opponent’s). A team-by-team analysis also highlighted both differences and similarities between the teams’ intrinsic dynamics, revealing common and unique team signatures -- that is, the way they adapted to this constraint. 

At this stage, the method used here is useful for comparing systems based on the patterns of interactions used, identifying how the systems adapt to a given constraint, and investigating the link with the functional outcome. There are many ways to improve the model and methodology presented here, but this work will contribute to the existing scientific literature on network analysis applied to real data, as it is easily applicable to other types of data.

\bibliographystyle{apalike}  
\bibliography{main}

\begin{thebibliography}{}

\bibitem[Aparicio et~al., 2016]{Aparicio2017}
Aparicio, D., Ribeiro, P., and Silva, F. (2016).
\newblock Extending the applicability of graphlets to directed networks.
\newblock {\em IEEE/ACM transactions on computational biology and bioinformatics}, 14(6):1302--1315.

\bibitem[Ara{\'u}jo et~al., 2022]{araujo2022team}
Ara{\'u}jo, D., Brito, H., and Carrilho, D. (2022).
\newblock Team decision-making behavior: An ecological dynamics approach.
\newblock {\em Asian Journal of Sport and Exercise Psychology}.

\bibitem[Balague et~al., 2013]{Balague2013}
Balague, N., Torrents, C., Hristovski, R., Davids, K., and Ara{\'u}jo, D. (2013).
\newblock Overview of complex systems in sport.
\newblock {\em Journal of Systems Science and Complexity}, 26:4--13.

\bibitem[Battiston et~al., 2020]{Battiston2020}
Battiston, F., Cencetti, G., Iacopini, I., Latora, V., Lucas, M., Patania, A., Young, J.-G., and Petri, G. (2020).
\newblock Networks beyond pairwise interactions: Structure and dynamics.
\newblock {\em Physics Reports}, 874:1--92.

\bibitem[Bekkers and Dabadghao, 2019]{bekkers2019flow}
Bekkers, J. and Dabadghao, S. (2019).
\newblock Flow motifs in soccer: What can passing behavior tell us?
\newblock {\em Journal of Sports Analytics}, 5(4):299--311.

\bibitem[Buldu et~al., 2019]{Buldu2019}
Buldu, J., Busquets, J., Echegoyen, I., and Seirul.~lo, F. (2019).
\newblock Defining a historic football team: Using network science to analyze guardiola’s fc barcelona.
\newblock {\em Scientific reports}, 9(1):13602.

\bibitem[C{\'a}rdenas et~al., 2015]{Cardenas2015}
C{\'a}rdenas, D., Ortega, E., Llorca, J., Courel, J., S{\'a}nchez-Delgado, G., and Isabel~Pi{\~n}ar, M. (2015).
\newblock Motor characteristics of fast break in high level basketball.
\newblock {\em Kinesiology}, 47(2.):208--214.
\newblock \url {https://hrcak.srce.hr/150548}.

\bibitem[Casteigts, 2018]{Casteigts2018}
Casteigts, A. (2018).
\newblock {\em {A Journey through Dynamic Networks (with Excursions)}}.
\newblock Habilitation {\`a} diriger des recherches, {Universit{\'e} de Bordeaux}.

\bibitem[Cintia et~al., 2016]{cintia2016haka}
Cintia, P., Coscia, M., and Pappalardo, L. (2016).
\newblock The haka network: Evaluating rugby team performance with dynamic graph analysis.
\newblock In {\em 2016 IEEE/ACM International Conference on Advances in Social Networks Analysis and Mining (ASONAM)}, pages 1095--1102.

\bibitem[Clemente et~al., 2015]{clemente2015network}
Clemente, F.~M., Martins, F. M.~L., Kalamaras, D., and Mendes, R.~S. (2015).
\newblock Network analysis in basketball: Inspecting the prominent players using centrality metrics.
\newblock {\em Journal of physical education and sport}, 15(2):212.

\bibitem[Correia et~al., 2013]{Correia2013}
Correia, V., Ara{\'u}jo, D., Vilar, L., and Davids, K. (2013).
\newblock From recording discrete actions to studying continuous goal-directed behaviours in team sports.
\newblock {\em Journal of sports sciences}, 31(5):546--553.

\bibitem[Courel-Ib{\'a}{\~n}ez et~al., 2017]{Courel-Ibanez2017}
Courel-Ib{\'a}{\~n}ez, J., McRobert, A.~P., Toro, E.~O., and V{\'e}lez, D.~C. (2017).
\newblock Collective behaviour in basketball: a systematic review.
\newblock {\em International Journal of Performance Analysis in Sport}, 17(1-2):44--64.

\bibitem[Duarte et~al., 2012]{Duarte2012}
Duarte, R., Ara{\'u}jo, D., Correia, V., and Davids, K. (2012).
\newblock Sports teams as superorganisms: Implications of sociobiological models of behaviour for research and practice in team sports performance analysis.
\newblock {\em Sports medicine}, 42:633--642.

\bibitem[Edelman and Gally, 2001]{edelman2001degeneracy}
Edelman, G.~M. and Gally, J.~A. (2001).
\newblock Degeneracy and complexity in biological systems.
\newblock {\em Proceedings of the National Academy of Sciences}, 98(24):13763--13768.

\bibitem[Fewell et~al., 2012]{Fewell2012a}
Fewell, J.~H., Armbruster, D., Ingraham, J., Petersen, A., and Waters, J.~S. (2012).
\newblock Basketball teams as strategic networks.
\newblock {\em PloS one}, 7(11):e47445.

\bibitem[Friston and Price, 2003]{friston2003degeneracy}
Friston, K.~J. and Price, C.~J. (2003).
\newblock Degeneracy and redundancy in cognitive anatomy.
\newblock {\em Trends in cognitive sciences}, 7(4):151--152.

\bibitem[Gama et~al., 2020]{Gama2020}
Gama, J., Dias, G., Passos, P., Couceiro, M., and Davids, K. (2020).
\newblock Homogeneous distribution of passing between players of a team predicts attempts to shoot at goal in association football: A case study with 10 matches.
\newblock {\em Nonlinear Dynamics, Psychology \& Life Sciences}, 24(3).
\newblock \url {https://shura.shu.ac.uk/id/eprint/26912}.

\bibitem[Holme, 2015]{Holme2015}
Holme, P. (2015).
\newblock Modern temporal network theory: a colloquium.
\newblock {\em The European Physical Journal B}, 88:1--30.

\bibitem[Holme and Saram{\"a}ki, 2012]{holme2012temporal}
Holme, P. and Saram{\"a}ki, J. (2012).
\newblock Temporal networks.
\newblock {\em Physics reports}, 519(3):97--125.

\bibitem[Hulovatyy et~al., 2015]{Hulovatyy2015}
Hulovatyy, Y., Chen, H., and Milenkovi{\'c}, T. (2015).
\newblock Exploring the structure and function of temporal networks with dynamic graphlets.
\newblock {\em Bioinformatics}, 31(12):i171--i180.

\bibitem[Korte, 2019]{korte2019network}
Korte, F. (2019).
\newblock {\em Network Analysis in Team Sports}.
\newblock PhD thesis, Technische Universit{\"a}t M{\"u}nchen.

\bibitem[Korte and Lames, 2018]{korte2018characterizing}
Korte, F. and Lames, M. (2018).
\newblock Characterizing different team sports using network analysis.
\newblock {\em Current Issues in Sport Science (CISS)}, 3:005--005.

\bibitem[Kostakis et~al., 2017]{kostakis2017discovering}
Kostakis, O., Tatti, N., and Gionis, A. (2017).
\newblock Discovering recurring activity in temporal networks.
\newblock {\em Data Mining and Knowledge Discovery}, 31(6):1840--1871.

\bibitem[Krejtz et~al., 2015]{krejtz2015gaze}
Krejtz, K., Duchowski, A., Szmidt, T., Krejtz, I., Gonz{\'a}lez~Perilli, F., Pires, A., Vilaro, A., and Villalobos, N. (2015).
\newblock Gaze transition entropy.
\newblock {\em ACM Transactions on Applied Perception (TAP)}, 13(1):1--20.

\bibitem[LaRock et~al., 2022]{LaRock2022}
LaRock, T., Scholtes, I., and Eliassi-Rad, T. (2022).
\newblock Sequential motifs in observed walks.
\newblock {\em Journal of Complex Networks}, 10(5):cnac036.

\bibitem[Malqui et~al., 2019]{malqui2019soccer}
Malqui, J. L.~S., Romero, N. M.~L., Garcia, R., Alemdar, H., and Comba, J.~L. (2019).
\newblock How do soccer teams coordinate consecutive passes? a visual analytics system for analysing the complexity of passing sequences using soccer flow motifs.
\newblock {\em Computers \& Graphics}, 84:122--133.

\bibitem[Mart{\'\i}nez et~al., 2020]{Martinez2020}
Mart{\'\i}nez, J.~H., Garrido, D., Herrera-Diestra, J.~L., Busquets, J., Sevilla-Escoboza, R., and Buld{\'u}, J.~M. (2020).
\newblock Spatial and temporal entropies in the spanish football league: A network science perspective.
\newblock {\em Entropy}, 22(2):172.

\bibitem[Martins et~al., 2020]{Martins2020}
Martins, F., Gomes, R., Lopes, V., Silva, F., and Mendes, R. (2020).
\newblock Node and network entropy—a novel mathematical model for pattern analysis of team sports behavior.
\newblock {\em Mathematics}, 8(9):1543.

\bibitem[Mason et~al., 2015]{mason2015hidden}
Mason, P.~H., Winter, B., Grignolio, A., et~al. (2015).
\newblock Hidden in plain view: degeneracy in complex systems.
\newblock {\em Biosystems}, 128:1--8.

\bibitem[Mattsson and Takes, 2021]{Mattsson2021}
Mattsson, C.~E. and Takes, F.~W. (2021).
\newblock Trajectories through temporal networks.
\newblock {\em Applied Network Science}, 6(1):35.

\bibitem[McCrum-Gardner, 2008]{mccrum2008correct}
McCrum-Gardner, E. (2008).
\newblock Which is the correct statistical test to use?
\newblock {\em British Journal of Oral and Maxillofacial Surgery}, 46(1):38--41.

\bibitem[McGarry et~al., 2002]{mcgarry2002sport}
McGarry, T., Anderson, D.~I., Wallace, S.~A., Hughes, M.~D., and Franks, I.~M. (2002).
\newblock Sport competition as a dynamical self-organizing system.
\newblock {\em Journal of sports sciences}, 20(10):771--781.

\bibitem[Meza, 2017]{meza2017flow}
Meza, D. A.~P. (2017).
\newblock Flow network motifs applied to soccer passing data.
\newblock In {\em Proceedings of MathSport international 2017 conference}, page 305.

\bibitem[Milo et~al., 2002]{Milo2002}
Milo, R., Shen-Orr, S., Itzkovitz, S., Kashtan, N., Chklovskii, D., and Alon, U. (2002).
\newblock Network motifs: simple building blocks of complex networks.
\newblock {\em Science}, 298(5594):824--827.

\bibitem[Moreno et~al., 2013]{moreno2013effects}
Moreno, E., G{\'o}mez, M.~A., Lago, C., and Sampaio, J. (2013).
\newblock Effects of starting quarter score, game location, and quality of opposition in quarter score in elite women’s basketball.
\newblock {\em Kinesiology}, 45(1.):48--54.
\newblock \url {https://hrcak.srce.hr/104547}.

\bibitem[Neuman et~al., 2018]{neuman2018adaptive}
Neuman, Y., Israeli, N., Vilenchik, D., and Cohen, Y. (2018).
\newblock The adaptive behavior of a soccer team: An entropy-based analysis.
\newblock {\em Entropy}, 20(10):758.

\bibitem[Newman, 2003]{newman2003structure}
Newman, M.~E. (2003).
\newblock The structure and function of complex networks.
\newblock {\em SIAM review}, 45(2):167--256.

\bibitem[Oberoi et~al., 2023]{Oberoi2023}
Oberoi, K.~S., Del~Mondo, G., Ga{\"u}z{\`e}re, B., Dupuis, Y., and Vasseur, P. (2023).
\newblock Detecting dynamic patterns in dynamic graphs using subgraph isomorphism.
\newblock {\em Pattern Analysis and Applications}, pages 1--17.

\bibitem[Paix{\~a}o et~al., 2015]{paixao2015does}
Paix{\~a}o, P., Sampaio, J., Almeida, C.~H., and Duarte, R. (2015).
\newblock How does match status affects the passing sequences of top-level european soccer teams?
\newblock {\em International Journal of Performance Analysis in Sport}, 15(1):229--240.

\bibitem[Paranjape et~al., 2017]{Paranjape2017}
Paranjape, A., Benson, A.~R., and Leskovec, J. (2017).
\newblock Motifs in temporal networks.
\newblock In {\em Proceedings of the tenth ACM international conference on web search and data mining}, pages 601--610.

\bibitem[Pereira et~al., 2021]{Pereira2021}
Pereira, L.~R., Lopes, R.~J., Lou{\c{c}}{\~a}, J., Ara{\'u}jo, D., and Ramos, J. (2021).
\newblock The soccer game, bit by bit: An information-theoretic analysis.
\newblock {\em Chaos, Solitons \& Fractals}, 152:111356.

\bibitem[Picciolo et~al., 2022]{Picciolo2022}
Picciolo, F., Ruzzenenti, F., Holme, P., and Mastrandrea, R. (2022).
\newblock Weighted network motifs as random walk patterns.
\newblock {\em New Journal of Physics}, 24(5):053056.

\bibitem[Pr{\v{z}}ulj et~al., 2004]{prvzulj2004modeling}
Pr{\v{z}}ulj, N., Corneil, D.~G., and Jurisica, I. (2004).
\newblock Modeling interactome: scale-free or geometric?
\newblock {\em Bioinformatics}, 20(18):3508--3515.

\bibitem[Ramos et~al., 2018]{Ramos2018}
Ramos, J., Lopes, R.~J., and Ara{\'u}jo, D. (2018).
\newblock What’s next in complex networks? capturing the concept of attacking play in invasive team sports.
\newblock {\em Sports medicine}, 48:17--28.

\bibitem[Sampaio et~al., 2010]{sampaio2010effects}
Sampaio, J., Lago, C., Casais, L., and Leite, N. (2010).
\newblock Effects of starting score-line, game location, and quality of opposition in basketball quarter score.
\newblock {\em European Journal of Sport Science}, 10(6):391--396.

\bibitem[Seifert et~al., 2017]{Seifert2017}
Seifert, L., Ara{\'u}jo, D., Komar, J., and Davids, K. (2017).
\newblock Understanding constraints on sport performance from the complexity sciences paradigm: An ecological dynamics framework.
\newblock {\em Human movement science}, 56:178--180.

\bibitem[Shannon, 1948]{Shannon1948}
Shannon, C.~E. (1948).
\newblock {A Mathematical Theory of Communication}.
\newblock {\em Bell System Technical Journal}, 27(3):379--423.

\bibitem[Skinner, 2010]{skinner2010price}
Skinner, B. (2010).
\newblock The price of anarchy in basketball.
\newblock {\em Journal of Quantitative Analysis in Sports}, 6(1).

\bibitem[{\v{S}}trumbelj et~al., 2013]{vstrumbelj2013decade}
{\v{S}}trumbelj, E., Vra{\v{c}}ar, P., Robnik-{\v{S}}ikonja, M., De{\v{z}}man, B., and Er{\v{c}}ulj, F. (2013).
\newblock A decade of euroleague basketball: An analysis of trends and recent rule change effects.
\newblock {\em Journal of human kinetics}, 38(2013):183--189.

\bibitem[Tononi et~al., 1999]{tononi1999measures}
Tononi, G., Sporns, O., and Edelman, G.~M. (1999).
\newblock Measures of degeneracy and redundancy in biological networks.
\newblock {\em Proceedings of the National Academy of Sciences}, 96(6):3257--3262.

\bibitem[Vernet et~al., 2023]{vernet2023study}
Vernet, M., Pign{\'e}, Y., and Sanlaville, E. (2023).
\newblock A study of connectivity on dynamic graphs: computing persistent connected components.
\newblock {\em 4OR}, 21(2):205--233.

\bibitem[Welch et~al., 2021]{Welch2021Collective}
Welch, M., Schaerf, T.~M., and Murphy, A. (2021).
\newblock Collective states and their transitions in football.
\newblock {\em Plos one}, 16(5):e0251970.

\bibitem[Whitacre, 2010]{whitacre2010degeneracy}
Whitacre, J.~M. (2010).
\newblock Degeneracy: a link between evolvability, robustness and complexity in biological systems.
\newblock {\em Theoretical Biology and Medical Modelling}, 7:1--17.

\bibitem[Xin et~al., 2017]{xin2017continuous}
Xin, L., Zhu, M., and Chipman, H. (2017).
\newblock A continuous-time stochastic block model for basketball networks.
\newblock {\em The Annals of Applied Statistics}, 11(2).

\bibitem[Yamamoto and Yokoyama, 2011]{yamamoto2011common}
Yamamoto, Y. and Yokoyama, K. (2011).
\newblock Common and unique network dynamics in football games.
\newblock {\em PloS one}, 6(12):e29638.

\bibitem[Zuccolotto et~al., 2018]{zuccolotto2018big}
Zuccolotto, P., Manisera, M., and Sandri, M. (2018).
\newblock Big data analytics for modeling scoring probability in basketball: The effect of shooting under high-pressure conditions.
\newblock {\em International journal of sports science \& coaching}, 13(4):569--589.

\end{thebibliography}

\newpage
\section*{Appendix}

\begin{algorithm}[H] 
    \caption{Algorithm running through all the possible classifications (with a filter)}\label{Algorithm_1}
    \begin{algorithmic}[1]
        \State $p \gets$ the minimum percentage of data a class must contain (i.e. the number of possessions in the class divided by the total number of possessions)
        \State $f \gets$ set of integers from the lowest relative score to the highest relative score
        \State $n \gets$ the length of $f$ (i.e. the number of values in $f$)
        \State $\text{c} \gets$ empty list of classifications
                \\
        \For{each team}
            \For{$f_{1}$ in $f[3]:f[n-3]$}
                \For{$f_{2}$ in $(f_{1}+2):f[n-1]$}
                    \State $class1 \gets f[1]:(f_{1}-1)$
                    \State $class2 \gets f_{1}:(f_{2}-1)$
                    \State $class3 \gets f_{2}:f[n]$
                    \If{$\text{percentage of data in } \text{$class1$, $class2$ and $class3$} \geq p$}
                        \State append the list $[class1,class2,class3]$ to $\text{c}$
                    \EndIf
                \EndFor
            \EndFor
        \EndFor
                \\
    \State \textbf{output:} $c$ contain all valid classifications
    \end{algorithmic}
\end{algorithm}

\end{document}